\documentclass[nofootinbib,twocolumn,aps,letterpaper,superscriptaddress,
]{revtex4}
\usepackage{amsmath}
\usepackage{amsfonts}
\usepackage{amssymb}
\usepackage{epsfig,graphicx,bm}
\usepackage{xcolor}

\usepackage{tikz-cd}
\usepackage{tikz}
\usepackage{lscape}
\usepackage{epstopdf}

\begin{document}

\newcommand{\mc}[1]{\mathcal{#1}}
\newcommand{\ms}[1]{\mathscr{#1}}
\newcommand{\mbb}[1]{\mathbb{#1}}
\newcommand{\mbf}[1]{\mathbf{#1}}
\newcommand{\mb}[1]{\boldsymbol{#1}}
\newcommand{\id}{\mathbb{1}}
\newcommand{\D}{\mathcal{D}}
\newcommand{\de}{\delta}
\newcommand{\p}{\partial}
\newcommand{\lm}{\lambda}
\newcommand{\eps}{\varepsilon}
\newcommand{\ket}[1]{|#1\rangle}
\newcommand{\Tr}{{\rm Tr}}
\newcommand{\w}{\wedge}
\newcommand{\hodge}{\phantom{}^*\!}
\newcommand{\ph}[1]{\phantom{#1}}
\newcommand{\mdots}{,.\,.\,,}
\newcommand{\margin}[1]{\marginpar{\scriptsize[ #1 ]}}
\newcommand{\be}{\nopagebreak[3]\begin{equation}}
\newcommand{\ee}{\end{equation}}
\newcommand{\bea}{\nopagebreak[3]\begin{eqnarray}}
\newcommand{\eea}{\end{eqnarray}}

\author{Niels Gresnigt}
\email{niels.gresnigt@xjtlu.edu.cn}
\affiliation{Department of Physics, Xi'an Jiaotong-Liverpool University, 215123 Suzhou, China}

\author{Antonino Marcian\`o}
\email{marciano@fudan.edu.cn}
\affiliation{Center for Field Theory and Particle Physics \& Department of Physics, Fudan University, 200433 Shanghai, China}
\affiliation{Laboratori Nazionali di Frascati INFN, Frascati (Rome), Italy, EU}

\author{Emanuele Zappala}
\email{emanuele.zappala@yale.edu, zae@usf.edu}
\affiliation{Yale School of Medicine, Yale University, New Haven, CT, USA}

\title{On the dynamical emergence of $SU_q(2)$ from the regularization of $2+1D$ gravity with cosmological constant}


\begin{abstract}
\noindent 
The quantization of the reduced phase-space of the Einstein-Hilbert action for gravity in $2+1D$ has been shown to bring about the emergence, at the quantum level, of a topological quantum field theory endowed with an $SU_q(2)$ quantum group symmetry structure. We hereby tackle the same problem, but start from the kinematical $SU(2)$ (quantum) Hilbert space of the theory of $2+1D$ gravity with non-zero cosmological constant in the Palatini formalism, and subsequently impose the constraints. We hence show the dynamical emergence of the $SU_q(2)$ quantum group at the quantum level within the spin-foam framework. 
The regularized curvature constraint is responsible for the effective representations of $SU_q(2)$ that are recovered for any Wilson loop evaluated at the $SU(2)$ group element that encodes the discretization of the space-time curvature induced by the cosmological constant. The extension to the spin-network basis, and consequently to any transition amplitude between its generic states, enables us to derive in full generality the recoupling theory of $SU_q(2)$. We provide constructive examples for the scalar product of two loop states and spin-networks encoding trivalent vertices. We further comment on the diffeomorphism symmetry generated by the implementation of the curvature constraint, and finally derive explicitly the partition function amplitude of the Turaev-Viro model. 
\end{abstract}


\maketitle

\section{Introduction}
\noindent 
The quantization of gravity in $2+1D$ is notoriously a solvable problem \cite{WittenQuantumGravity2+1D}, as it has been emphasised over the last four decades in the literature on topological quantum field theory. There are two fundamental directions to achieve this goal in reduced space-time dimensions. Focusing on the Euclidean $SU(2)$ symmetric version of $2+1D$ gravity, it is possible either to accomplish the quantization of the reduced phase-space of the classical theory, imposing classical constraints before quantization \cite{WittenQuantumGravity2+1D}, or to achieve directly the quantization of the kinematical Hilbert space of the theory, on which only the gauge constraint has been imposed.  

Having the cosmological constant involved in the analysis allows for a novel symmetric structure to arise, expressed in terms of the axioms of Hopf algebras. These axioms in turn can be cast resorting to the versatility of the Reidemeister moves \cite{Kauffman:1990am, Kauffman1994}. Jones polynomials are crucial in recovering the link to quantum Hopf algebras, and to show how these latter ones provide link invariants via solutions to the Yang-Baxter equations. 
Relevant topological invariants, including the Jones polynomials \cite{WittenJonesPolinomials}, are defined in terms of their properties under Reidemeister moves. This in general characterises a wide field of studies, encoding topological quantum field theory.

In particular, the novel symmetric structures that emerge 
are called quantum groups. These are non-trivial Hopf algebras, characterised by the deformation of the product rules in the algebraic sector and the deformation of the Leibnitz rule at the level of the co-algebra, which in turn enters the bi-algebra structure of Hopf algebras. 
On the side of the quantization of the reduced phase-space of gravity in $2+1D$ with cosmological constant, it was proven by Witten \cite{WittenJonesPolinomials} that the path-integral quantization of the theory, equivalent to the quantization of two uncoupled Chern-Simons theories, provides the Turaev-Viro topological invariant \cite{TV}. Nonetheless, it has remained unclear hitherto whether a different procedure of quantization, accounting for the imposition at the quantum level of the curvature constraint on the states of the kinematical Hilbert space, would entail the same theory.

Although several other authors have likewise considered the canonical quantization of $2+1D$ gravity with a cosmological constant, the approach taken here differs substantially from earlier works. The novelty of our study lies in the fact that we have shown that recoupling theory of quantum groups appears at the quantum level as a by-product of the quantization procedure, due to the implementation of the quantum constraints at the dynamical level, using the undeformed phase-space of the BF formulation of gravity with cosmological constant in $2+1D$. 
In Refs.~\cite{Perez:2010pm,Noui:2011im,Noui:2011aa}, a grasping procedure was implemented on the holonomies in order to derive the quantum group recoupling theory in a theory of gravity with cosmological constant. Ref.~\cite{Pranzetti:2014xva} utilized the connections of the two Chern-Simons theories in which gravity can be reformulated. This solution to the problem still leaves unanswered the link between the quantization of the Palatini formulation of gravity with cosmological constant and the Turaev-Viro model. As far as the present work is concerned, we moved from the issue raised by Witten in the 80's \cite{WittenQuantumGravity2+1D}, in turn  elaborating on related work by Atiyah and Bott \cite{Atiyah:1982fa}, of whether or not it is possible to derive the quantum group recoupling theory for gravity with nontrivial cosmological constant by imposing constraints at the quantum level. We tackled this question from the Palatini/BF-like formulation of gravity, without invoking the Chern-Simons theory, finding a positive answer.

As an advantage, our procedure does not require any restriction on either the sign of the cosmological constant, nor on its (discretized) value, and hence holds in full generality. Other studies that have considered the aforementioned perspective suggested in \cite{WittenQuantumGravity2+1D}, although following different approaches, are: Ref.~\cite{Bonzom:2014bua}, which introduced recoupling theory of quantum groups based on an algebraically motivated Poisson-Lie deformation procedure of the discretized topological BF theories developed in \cite{Bonzom:2014wva}; Ref.~\cite{Dupuis:2020ndx}, which used a canonical transformation that deforms the gauge invariance and the boundary symmetries of the theory, introducing a dependence on the cosmological constant, and realised a discretization procedure that induces a truncation of the degrees of freedom in the continuum.

The plan of the paper is the following. In section \ref{can3D} we review the canonical Palatini formulation of gravity with cosmological constant in $3D$. Section \ref{SU2kin} introduces the quantization of the theory at the level of the kinematical Hilbert space, without imposing yet the constraints. In section \ref{quantumsym} we comment on the physical relevance of the quantum symmetrizer and prove that the quantum dynamics of the $SU(2)$ theory induces the quantum recoupling of the Turaev-Viro model. In section \ref{sf} we recall the spin-foam dynamics formalism, complemented with cosmological constant, and then we extend the Noui-Perez physical projector so to encode the cosmological constant in the Palatini formalism. In section \ref{reg} we perform the regularization of the curvature constraint of the Palatini formulation of gravity with cosmological constant. In section \ref{2l} we apply the extended and regularized physical projector to the study-case of the scalar product of two loops. Section \ref{diffeo} deals with the diffeomorphism invariance of the extended and regularized physical projector. Finally, section \ref{tuvi} explicitly shows the emergence of the Turaev-Viro model from the classical theory started from.
In section \ref{con} we spell out our conclusions.

\section{Canonical 3D gravity with cosmological constant}\label{can3D}
\noindent 
The (first order formalism) three-dimensional Riemannian theory of gravity with cosmological 
constant $\Lambda$ that we are considering is defined on a space-time $\mathcal{M}$, which we assume to be a three-dimensional oriented smooth manifold, through the expression for the action 
\be \label{ac}
S[e,\omega]=\int_{\mathcal{M}} {\rm Tr}[{e} \wedge F({\omega})] + \frac{\Lambda}{3} \,{\rm Tr}[e \wedge e \wedge e]\,,
\ee
where $e$ stands for the triad, which is an $\mathfrak{su}(2)$-valued $1$-form, $\omega$ is an $SU(2)$ three dimensional connection, $F(\omega)$ is the curvature of $\omega$ and the trace ``Tr'' denotes the Killing form on $\mathfrak{su}(2)$. With no loss of generality, we can adopt the usual decomposition and assume the space-time topology to be $\mathcal{M} = \Sigma \times \mathbb{R}$, where $\Sigma$ is a Riemann surface of arbitrary genus.

Suppose now to pull back to $\Sigma$ the spin-connection $\omega$ and the triad $e$, then we can express the new variables in the local coordinates to be the two-dimensional connection $A^i_a$ and the triad field $e_b^j$, in which $a=1,2$ are space coordinate indices and $i,\,j=1,\,2,\,3$ are $\mathfrak{su}(2)$ indices. The Poisson brackets among these variables now provide the symplectic structure
\be
\{ A^i_a(x),\,e_b^j(y)\}=\epsilon_{ab}\,\delta^{ij} \, \delta^{(2)}(x,y)\,.
\ee
The phase space of the theory can also be parametrized in terms of the desitized triad $E^b_j=\epsilon^{bc}\,e^k_c\,\eta_{jk}$, {\it i.e.}
\be
\{ A^i_a(x),\,E^b_j(y)\}=\delta^b_a\,\delta^i_j \delta^{(2)}(x,y)\,.
\ee
Varying the action in terms of the pull back of the fields, {\it i.e.} with respect to the independent fields $A^i_a(x)$ and $e_b^j(x)$, we get the first class local constraints $D_A\,e\simeq 0$ --- here $\simeq$ denotes validity on the constraints surface, namely weak equality --- and $F(A)+ \Lambda e \wedge e \simeq 0$. In terms of the components, we find  
\be
D_A^b\,e_b^j=0\,,  \qquad
F^i_{ab}(A)+ \Lambda \epsilon^{i}_{\;jk} e_a^j\,e_b^k = 0\,.
\ee
These constraints generate local symmetries. In particular, smearing out $D_A^b\,e_b^j$ with the test field $\alpha_j$ we get the Gau\ss\, constraint
\be
G[\alpha,\, A,\, e]=\int_{\Sigma} \alpha_j\, D_A^b\,e_b^j =0\,,
\ee
which generate infinitesimal $SU(2)$ gauge transformations
\bea
\delta_\alpha A^i_a&=&\{ A^a_i,\, G[\alpha,\, A,\, e]\} = (D_a\alpha)^i\,, \nonumber\\
\delta_\alpha e_a^i&=&\{ e_a^i,\, G[\alpha,\, A,\, e]\} = \alpha_k e_{aj} \epsilon^{ijk}\,.
\eea
Smearing out $F^i_{ab}(A)+ \Lambda \epsilon^{i}_{\;jk} e_a^j\,e_b^k$ with the test function $\beta_j$, we get the curvature constraint $C_\Lambda[\beta, A]$, which reads
\be
C_\Lambda[\beta, A,e]=\int_{\Sigma} \beta_i\,(F^i_{ab}(A)+ \Lambda \epsilon^{i}_{\;jk} e_a^j\,e_b^k)=0 
\ee
and generates transformations that contain diffeomorphisms, namely
\bea
\delta_\beta A^i_a&=&\{ A^i_a,\,C_\Lambda[\beta, A,e]  \}= \Lambda \epsilon^{ijk} e^j_a \beta^k\,, \nonumber\\
\delta_\beta e^i_a&=&\{ e^i_a,\,C_\Lambda[\beta, A,e]  \}=D_c \beta^i\,,
\eea
provided that the triad fields $e^i_a$ are assumed to be non degenerate.

Indeed, if we consider the vector field $v=v^a\partial_a$ on the surface $\Sigma$ and hence define the parameters $\alpha^i=v^a\,A^i_a$ and $\beta^i=e_a^i \,v^b$, the previous transformations become\footnote{We recall that spatial diffeomorphism along a vector field $v^a$ are defined by $\delta_v A^i_a=\{ A^i_a, V(v^a) \}=\mathcal{L}_v A^i_a$ and $\delta_v e^i_a=\{ e^i_a, V(v^a) \}=\mathcal{L}_v e^i_a$, where $V$ is the canonical vector constraint of general relativity.}
\be
(\mathcal{L}_vA)^i_a \simeq \delta_{\alpha(v)} A_a^i\,,  \quad (\mathcal{L}_v e)^i_a \simeq  \delta_{\alpha(v)} e^i_a +  \delta_{\beta(v)} e^i_a\,,
\ee
where $\mathcal{L}_v$ is the Lie derivative along the vector field $v$.

\section{$SU(2)$ kinematical Hilbert space. } \label{SU2kin}
\noindent 
The theory above can be quantized {\it \`a la loop} by a way \cite{Rov, Thi} that follows the Dirac's procedure. Indeed, we can first construct an auxiliary Hilbert space on which we provide a representation of the basic variables we are going to deal with and on which constraints will be represented. In our scheme, connections are represented in terms of holonomies $h_\gamma[A]$ along path $\gamma\in \Sigma$, that are in turn defined by $h_\gamma[A]=P \exp \int_\gamma A$, where $P$ denotes here path ordering. Thus functional of connections will be represented in terms of functionals of holonomies. Triad fields $e^i_a$, associated to the densitized electric field $E_i^a$, will be smeared, as usual, along co-dimension one surfaces. These canonical variables are then promoted to operators acting on the auxiliary Hilbert space of functionals of holonomies. The physical Hilbert space corresponds to those states that are annihilated by the constraints. These states are distributional, as they are not normalizable with respect to the auxiliary Hilbert space and hence no more in it.

The auxiliary Hilbert space $\mathcal{H}_{aux}$ is the Cauchy completion of the space of cylindrical functions $Cyl$. These latter ones are defined on the space of generalized connections $\mathcal{A}$, which provide in turn a map from the set of paths $\gamma\in \Sigma$ to $SU(2)$, and hence represent an extension of the notion of holonomy $h_\gamma[A]$. Elements $\Psi_{\Gamma,f}[A]$ of the space $Cyl$ are defined as follows
\be
\Psi_{\Gamma,f}[A]= f(h_{\gamma_1}[A],\cdot \cdot\cdot h_{\gamma_l}[A],\cdot \cdot\cdot h_{\gamma_L}[A] )\,.
\ee
Consequently, these states are functionals of $A$ labeled by a finite graph $\Gamma\in \Sigma$ and a continuous function $f: \, SU(2)^L \rightarrow{} \mathbb{C}$, where $L$ denotes the number of links $\gamma_l$ of $\Gamma$. 
Therefore, the inner product adopted in order to define the completion of the auxiliary Hilbert space is that one used for any two cylindrical functions $\Psi_{\Gamma_1,f_1}[A]$ and $\Psi_{\Gamma_2,f_2}[A]$, namely the Ashtekar-Lewandowski measure 
\bea \label{Ash-Lew}
&\!\!\!\!\!\!\!\!\!\!\!\!\!\!\!\!\!\!\phantom{a}\!\!&\mu(\overline{\Psi_{\Gamma_1,f_1}[A]}\Psi_{\Gamma_2,f_2}[A])\equiv <\Psi_{\Gamma_1,f_1}[A],\, \Psi_{\Gamma_2,f_2}[A] >=\nonumber\\
&=&\int \prod \limits_{\tilde{l}=1}^{\tilde{L}} dh_{\tilde{l}} \, \overline{\tilde{f}_1(\cdot \cdot\cdot h_{\gamma_{\tilde{l}}}[A],\cdot \cdot\cdot h_{\gamma_{\tilde{L}}}[A] )}\cdot \nonumber\\  &\,\,\phantom{a}\,\,&\cdot \tilde{f}_2(\cdot \cdot\cdot h_{\gamma_{\tilde{l}}}[A],\cdot \cdot\cdot h_{\gamma_{\tilde{L}}}[A] )\,, \nonumber
\eea
in which $\tilde{l}$ labels links of $\tilde{\Gamma}=\Gamma_1 \cup \Gamma_2 $ (whose total number of links is $\tilde{L}$) and $\tilde{f}_1$ and $\tilde{f}_2$ denote the extension of the functions $f_1$ and $f_2$, defined respectively on $\Gamma_1$ and $\Gamma_2$, on $\tilde{\Gamma}$, and $dh_{\tilde{l}}$ stands for the invariant $SU(2)$-Haar measure.

Quantization on $\mathcal{H}_{aux}$ of generalized connections is achieved by promoting holonomies to act as operators on $\mathcal{H}_{aux}$, namely
\be
\widehat{h_\gamma[A]}\,\Psi[A]=h_\gamma[A]\, \Psi[A]\,,
\ee
whose procedure defines a self adjoint operator in $\mathcal{H}_{aux}$. In a similar way, the triad $e^i_a$ is promoted to a self adjoint operator valued distribution acting as a derivative with respect to $A$, {\it i.e.}
\be
\widehat{e}_i^a= - i\,L_P \, \frac{\partial}{\partial A^i_a}\,,
\ee
and equivalently the densitized Ashtekar electric field becomes
\be
\widehat{E}^i_a= - i\,L_P \epsilon_{ab} \eta^{ij} \frac{\partial}{\partial A^j_b}\,,
\ee
in which $L_P=\hbar G$ ($G$ being the Newton constant) is the Planck length in three dimensions.

Imposition of the Gau\ss\, constraint corresponds to the selection of elements of $Cyl$ invariant under $SU(2)$ gauge transformations. Concretely, gauge transformations act 
on the cylindrical functions by acting on the holonomies as
\be
h_l[A] \rightarrow{} g_{s(l)} \,h_l[A]\, g_{t(l)}^{-1}\,,
\ee
in which $g_{s(l)},\,g_{t(l)}\in SU(2)$ are group elements associated, respectively, to the source and target nodes of the link $l$. The kernel of the Gau\ss\, constraint, namely the projection into the $SU(2)$ gauge invariant subspace of the auxiliary Hilbert space, defines the kinematical Hilbert space $\mathcal{H}_{kin} \subset \mathcal{H}_{aux}$. 

Harmonic analysis on $SU(2)$, and specifically the Peter-Weyl theorem, enables us to expand any square integrable function $f: SU(2) \rightarrow \mathbb{C}$ in terms of unitary irreducible representations of $SU(2)$ 
\be
f(h)=\sum_j f_j \stackrel{j}{\Pi}(h)\,, \quad {\rm with} \quad f_j=\int dh \overline{ \stackrel{j}{\Pi}(h)}\, f(h)\,,
\ee
in which $f_j$ can be seen as an element of the tensor product vector space $ \mathcal{H}_j^* \otimes \mathcal{H}_j $ (where $ \mathcal{H}_j$ denotes the vector space in the $j$ representation and $ \mathcal{H}_j^*$ represents its complex conjugated copy), and magnetic indices contraction is understood.

This procedure clearly enables to introduce an orthonormal basis of states in $\mathcal{H}_{aux}$. Any element of $Cyl$ can be now expressed as a linear combination of tensor product of $L$ $SU(2)$-irreducible representations. Orthogonality of such elements of $Cyl$ is checked by using the physical inner-product (\ref{Ash-Lew}). The action of the $SU(2)$ gauge transformations generator, {\it i.e.} the Gau\ss\, constraint,  on Fourier modes is given by 
\be
\stackrel{j}{\Pi}(h) \rightarrow{} \stackrel{j}{\Pi}(g_{s(l)}) \, \stackrel{j}{\Pi}(h_l)\, \stackrel{j}{\Pi}(g_{t(l)}^{-1})\,,
\ee
and allows to construct a basis of gauge invariant functions by contraction of Wigner representation matrices with $\mathfrak{su}(2)$-invariant tensor or $\mathfrak{su}(2)$-intertwiners.  Intertwiners that are $\mathfrak{su}(2)$-invariant admit an orthomormal basis $\iota_n\in {\rm Inv}[\mathcal{H}_{j_1}\otimes\mathcal{H}_{j_2}\otimes \cdot \cdot \cdot \otimes \mathcal{H}_{j_L}]$ and are labeled by $n$. Once we have introduced such a notation, we are able to define a basis of gauge-invariant elements of $Cyl$ that corresponds to the spin-network basis, whose element are labeled by a graph $\Gamma$, a set of spin $\{j_l\}$ for each link $l\in \Gamma$ and a set of intertwiners $\iota_n$ labeling nodes $n\in \Gamma$ 
\be
\psi_{\Gamma,\{j_l\}, \{\iota_n\} }[A]= \bigotimes_{n \in \Gamma}  \iota_n\, \bigotimes_{l \in \Gamma}\, \stackrel{j_{\gamma}}{\Pi}(h_l [A]) \,.
\ee

\section{The quantum symmetrizer in the loop basis} \label{quantumsym}
\noindent 
Historically, the loop basis was introduced by Rovelli and Smolin in \cite{RS1,RS2}, in order to implement the Wilsonian quantization of the Einstein-Hilbert theory of gravity, recast in term of the gauge ``Ashtekar'' variables \cite{Ash}. The elements of this basis are the Wilson loops, which are traces of closed holonomies of the gravitational $SU(2)$ gauge connection. These are automatically gauge invariant due to the properties of traces. It was then shown in \cite{Rovelli:1995ac} that the same basis is equivalent to the spin-network basis. Roughly speaking, the equivalence of the two bases is obtained by undoing the symmetrizer at the edges of given spin-network basis elements, therefore projecting each bundle of strands onto the Temperley-Lieb algebra with the same number of strands. This same principle will be used below as well, as we shall see. 
Loops introduced in this way are kinematical objects --- they do not account for the imposition of the space diffeomorphism and time re-parametrization constraints that implement the dynamics of the Einstein-Hilbert action --- and apply to the quantization of the Hilbert space of the $2+1D$ reduction of the same theory, which reduces to a topological quantum field theory.

The building block for our considerations is the loop in the fundamental representation (rep) of $SU(2)$. This is expressed by the Wigner matrix ${\Pi}(g)$, which provides the fundamental representation of an element $g\in SU(2)$. When the dependence on the group element is frozen by setting it equal to the unit element $g=e$, the trace of the Wigner matrix, namely the Wilson loop of the fundamental rep, provides the dimension of this latter one.

We may introduce for convenience graphical notations for the irreducible representations (irreps) of $SU(2)$. Spinor indices are denoted as $A,B\in\{0,1\}$. The trivial (identity) intertwiner $\delta^{\ A}_{B}$ is graphically denoted as a straight line, while the Levi-Civita tensors $\epsilon_{AB}$ and $\epsilon^{AB}$ are denoted respectively as bottom-up and bottom-down arches. Within this notation, a Wilson loop reads
\be \label{s2}
\begin{tikzpicture}[baseline={([yshift=-0.1cm]current bounding box.center)},vertex/.style={anchor=base,
circle,fill=black!25,minimum size=18pt,inner sep=2pt}]
\draw (0,0) circle (15pt);
\end{tikzpicture}
= 
\delta^A_B \stackrel{\frac{1}{2}}\Pi\!\!\!\!\!\! \phantom{\Pi}^B_A \, (g) |_{g=e}\equiv {\chi}_{\frac{1}{2}}(g)\,, 
\ee
where $\chi(g)$ denotes the character of the irrep ${\Pi}(g)$ of a group element $g\in SU(2)$. Furthermore, taking into account this diagrammatics, we can construct the Jones-Wenzl projector moving from the realization of the symmetrizer for two fundamental reps, so as to realize an irrep of spin $1$. The coefficients at the right hand-side of Eq.~\eqref{s2} are determined by the properties of the Wigner matrices (instantiating the irreps of $SU(2)$) with $g=e$, once the symmetrizer of the two fundamental reps is required to be a projector --- by iteration for $n>2$ fundamental reps, the Jones-Wenzl projector for a generic irrep of spin $j=n/2$ is recovered.

We observe that setting $g\!=\!e$ is equivalent to imposing the curvature flatness condition on the elements of the kinematical Hilbert space of the theory that is considered. In particular, in $2+1D$ this corresponds to imposing the curvature constraint of the Einstein-Hilbert action. Technically, this is realized by considering the action of the Dirac delta function $\delta(g)$ on loops, or equivalently on holonomies. The dimension of the fundamental representation of $SU(2)$ reads in this case

\be \label{abc}
d= \delta(g ) \triangleright \chi_{\frac{1}{2}}(g) \,,
\ee
where the action $\triangleright$ of $\delta(g )$ is normalized by integration with respect to the Haar measure $dg$. From now on, we focus on the $2+1D$ Einstein-Hilbert action with cosmological constant $\Lambda$. The imposition of the curvature constraint to the loops, in the fundamental representation, entails to calculate the trace of the holonomy of the Ashtekar connection in a $SU(2)$ group element $H_{\Lambda}$ that represents the space-time curvature induced by $\Lambda$.

The action of the curvature constraint when a non-vanishing $\Lambda$ is added to the Einstein-Hilbert theory can be derived applying a discretization procedure. Consequently the curvature group element $H_{\Lambda}$ can be written as in Eq.~\eqref{eq:HLambda}, where the dependence on the square root of the cosmological constant is specified --- see section \ref{reg} for details. At the quantum level, the curvature constraint amounts to the multiplication of the Hilbert space elements by the Dirac delta function $\delta(g\, H_{\Lambda}^{-1})$, namely
\begin{eqnarray}
	\begin{tikzpicture}[baseline={([yshift=-0.3cm]current bounding box.center)},vertex/.style={anchor=base,
		circle,fill=black!25,minimum size=18pt,inner sep=2pt}]
	\draw (0,0) circle (15pt);
	\draw[fill=black] (-0.52,0) circle (2pt);
	\node (a)  at (0.5,0.7) {$\frac{1}{2}$};
	\end{tikzpicture}
	&=& \delta(gH_\Lambda^{-1})\chi^{\frac{1}{2}}(g) = \stackrel{\frac{1}{2}}\Pi\!\!\!\!\!\! \phantom{\Pi}^{\alpha}_{\alpha} (gH_\Lambda^{-1}) \,,\label{eq:diag_loop} \\
\chi^{\frac{1}{2}}(H_\Lambda) &=& {\rm Tr}_{\frac{1}{2}}(e^{\sqrt{\Lambda}\tau_3n^3}) = 2 \cos({\sqrt{\Lambda}n^3}/{2})  \,,\label{eq:trace_loop}
\end{eqnarray}
with $\tau_3$ an anti-Hermitian basis element of $\mathfrak{su}(2)$ in the fundamental representation.
This immediately provides the fundamental representation 
\be 
d_q= {\chi}_{\frac{1}{2}} (H_{\Lambda})
\ee
of $SU_q(2)$, i.e. the Chebyschev polynomial of degree one in which the parameter $q$ is a function of $\sqrt{\Lambda}$. Projecting two strands on the loop basis, i.e. projecting two open strands on the Temperley-Lieb algebra, along with the value of the trace just computed provides the Jones-Wenzl projector at $q\neq -1$. In fact, in the Temperley-Lieb algebra we have

	\begin{eqnarray}\label{eq:qJW_2_strings}
	\begin{tikzpicture}[scale=0.4,baseline={([yshift=-0.3cm]current bounding box.center)},vertex/.style={anchor=base,
		circle,fill=black!25,minimum size=18pt,inner sep=2pt}]
	\draw (-1,-2) -- (-1,0);
	\draw (1,-2) -- (1,0);
	\draw (-1.5,0) rectangle (1.5,1);
	\draw (-1,1) -- (-1,3);
	\draw (1,1) -- (1,3); 
	\draw[fill=black] (1,-1) circle (4pt);
	\draw[fill=black] (-1,-1) circle (4pt);
	\end{tikzpicture}
	&=& 
 a\ \ 
 \begin{tikzpicture}[scale=0.4,baseline={([yshift=-0.3cm]current bounding box.center)},vertex/.style={anchor=base,
 	circle,fill=black!25,minimum size=18pt,inner sep=2pt}]
 \draw (-1,-2) -- (-1,3);
 \draw (1,-2) -- (1,3);
 \draw[fill=black] (-1,-1) circle (4pt);
 \draw[fill=black] (1,-1) circle (4pt);
 \end{tikzpicture}
 + 
 b \ \ 
 \begin{tikzpicture}[scale=0.4,baseline={([yshift=-0.3cm]current bounding box.center)},vertex/.style={anchor=base,
 	circle,fill=black!25,minimum size=18pt,inner sep=2pt}]
 \draw (-1,2) arc (180:360:1.5cm and 1.5cm); 
 \draw (2,-2) arc (0:180:1.5cm and 1.5cm); 
 \draw[fill=black] (-0.75,-1.2) circle (4pt);
 \draw[fill=black] (1.75,1.2) circle (4pt);
 \end{tikzpicture} \,.
	\end{eqnarray}
Imposing that the symmetrizer is a projector ({\it i.e.} it is idempotent) fixes the coefficients to be $a=1$ and $b=-d_q^{-1}$, which shows the claim. 
We consider now the symmetrizer on a number of strands larger than two, showing an iterative version of the reasoning given in Eq.~\eqref{eq:qJW_2_strings}. In fact, using the symmetrizer to project on Temperley-Lieb algebra elements, along with the value of the ``fundamental loop'' computed in Eq.~\eqref{eq:diag_loop} and Eq.~\eqref{eq:trace_loop}, automatically forces the Kauffman smoothing relations where the smoothing factor is provided by the quantum dimension. This is essentially a consequence of the fact that the Kauffman bracket is unique, and the value as computed in Eq.~\eqref{eq:trace_loop} fixes the value of the coefficients to be the quantum dimension. We want to prove that using the symmetrizer with group element corresponding to the cosmological constant necessarily satisfies the following inductive equation, found at the end of Section~3.2 of \cite{KL}, which shows that this is the Jones-Wenzl projector with arbitrary $q$:
\begin{eqnarray}
    \begin{tikzpicture}[baseline={([yshift=-0.1cm]current bounding box.center)},vertex/.style={anchor=base,
    	circle,fill=black!25,minimum size=18pt,inner sep=2pt}]]
        \draw (0,2) -- (0,1);
        \draw (0.2,2) -- (0.2,1);
        \draw[dashed] (0.25,1.5) -- (0.95,1.5);
        \draw (1,2) -- (1,1);
        \draw (-0.2,1) rectangle (1.2,0.5);
        \draw (0,0.5) -- (0,-0.5);
        \draw (0.2,0.5) -- (0.2,-0.5);
        \draw[dashed] (0.25,0) -- (0.95,0);
        \draw (1,0.5) -- (1,-0.5);
    \end{tikzpicture}
    &=& \label{eq:higher_smoothing}
  \ \  \begin{tikzpicture}[baseline={([yshift=-0.1cm]current bounding box.center)},vertex/.style={anchor=base,
    	circle,fill=black!25,minimum size=18pt,inner sep=2pt}]]
        \draw (0,2) -- (0,1);
        \draw (0.2,2) -- (0.2,1);
        \draw[dashed] (0.25,1.5) -- (0.9,1.5);
        \draw (0.95,2) -- (0.95,1);
        \draw (1.3,2) -- (1.3,-0.5);
        \draw (-0.2,1) rectangle (1.15,0.5);
        \draw (0,0.5) -- (0,-0.5);
        \draw (0.2,0.5) -- (0.2,-0.5);
        \draw (0.95,0.5) -- (0.95,-0.5);
        \draw[dashed] (0.25,0) -- (0.9,0);
    \end{tikzpicture}
    -
     \mu_n\ \ 
    \begin{tikzpicture}[baseline={([yshift=-0.2cm]current bounding box.center)},vertex/.style={anchor=base,
    	circle,fill=black!25,minimum size=18pt,inner sep=2pt}]]
    \draw (0,2) -- (0,1.7);
    \draw (0.2,2) -- (0.2,1.7);
    \draw (0.95,2) -- (0.95,1.7);
    
    \draw[dashed] (0.3,1.9) -- (0.9,1.9);
    
    \draw (-0.2,1.7) rectangle (1.15,1.2);
    
        \draw (0,1.2) -- (0,0.3);
        \draw (0.2,1.2) -- (0.2,0.3);
        
        \draw[dashed] (0.3,0.75) -- (0.9,0.75);
    
    \draw (-0.2,0.3) rectangle (1.15,-0.2);
    
    \draw (0,-0.2) -- (0,-0.7);
    \draw (0.2,-0.2) -- (0.2,-0.7);
    \draw[dashed] (0.3,-0.45) -- (0.95,-0.45);
    \draw (0.95,-0.2) -- (0.95,-0.7);

     \draw[rounded corners] (1.2,0.8) ..controls(1.4,1).. (1.5,2);
     \draw[rounded corners] (0.95,1.2) .. controls(0.95,0.85).. (1.2,0.8);
     
    \draw[rounded corners] (1.2,0.6) ..controls(1.4,0.8).. (1.5,-0.7);
     \draw[rounded corners] (0.95,0.3) .. controls(1.05,0.5).. (1.2,0.6);
    \end{tikzpicture}\,,
\end{eqnarray}
where $\mu_1 := \frac{1}{d_q}$ and $\mu_{n+1} = \frac{1}{d_q - \mu_n}$.
We proceed inductively, using as base for induction the result already displayed for two strands. Suppose that the statement holds for some $k>2$. Applying the symmetrizer on $k+1$ strands we find two types of elements of the Temperley-Lieb algebra. Those where the $(k+1)^{\rm th}$-strand is straight, and those where it is not. We depict this situation diagrammatically as
\begin{eqnarray}
    \begin{tikzpicture}[baseline={([yshift=-0.1cm]current bounding box.center)},vertex/.style={anchor=base,
    	circle,fill=black!25,minimum size=18pt,inner sep=2pt}]]
        \draw (0,2) -- (0,1);
        \draw (0.2,2) -- (0.2,1);
        \draw[dashed] (0.25,1.5) -- (0.95,1.5);
        \draw (1,2) -- (1,1);
        \draw (-0.2,1) rectangle (1.2,0.5);
        \draw (0,0.5) -- (0,-0.5);
        \draw (0.2,0.5) -- (0.2,-0.5);
        \draw[dashed] (0.25,0) -- (0.95,0);
        \draw (1,0.5) -- (1,-0.5);
        \draw[fill=black] (0,0) circle (2pt);
        \draw[fill=black] (0.2,0) circle (2pt);
        \draw[fill=black] (1,0) circle (2pt);
    \end{tikzpicture}
    &=& \label{eq:higher_smoothing}
  A\ \  \begin{tikzpicture}[baseline={([yshift=-0.1cm]current bounding box.center)},vertex/.style={anchor=base,
    	circle,fill=black!25,minimum size=18pt,inner sep=2pt}]]
        \draw (0,2) -- (0,1);
        \draw (0.2,2) -- (0.2,1);
        \draw[dashed] (0.25,1.5) -- (0.9,1.5);
        \draw (0.95,2) -- (0.95,1);
        \draw (1.3,2) -- (1.3,-0.5);
        \draw (-0.2,1) rectangle (1.15,0.5);
        \draw (0,0.5) -- (0,-0.5);
        \draw (0.2,0.5) -- (0.2,-0.5);
        \draw (0.95,0.5) -- (0.95,-0.5);
        \draw[dashed] (0.25,0) -- (0.9,0);
        \draw[fill=black] (0,0) circle (2pt);
        \draw[fill=black] (0.2,0) circle (2pt);
        \draw[fill=black] (1.3,0) circle (2pt);
        \draw[fill=black] (0.95,0) circle (2pt);
    \end{tikzpicture}
    +
    B\ \ 
    \begin{tikzpicture}[baseline={([yshift=-0.2cm]current bounding box.center)},vertex/.style={anchor=base,
    	circle,fill=black!25,minimum size=18pt,inner sep=2pt}]]
    \draw (0,2) -- (0,1.7);
    \draw (0.2,2) -- (0.2,1.7);
    \draw (0.95,2) -- (0.95,1.7);
    
    \draw[dashed] (0.3,1.9) -- (0.9,1.9);
    
    \draw (-0.2,1.7) rectangle (1.15,1.2);
    
        \draw (0,1.2) -- (0,0.3);
        \draw (0.2,1.2) -- (0.2,0.3);
        
        \draw[dashed] (0.3,0.75) -- (0.9,0.75);
    
    \draw (-0.2,0.3) rectangle (1.15,-0.2);
    
    \draw (0,-0.2) -- (0,-0.7);
    \draw (0.2,-0.2) -- (0.2,-0.7);
    \draw[dashed] (0.3,-0.45) -- (0.95,-0.45);
    \draw (0.95,-0.2) -- (0.95,-0.7);

     \draw[rounded corners] (1.2,0.8) ..controls(1.4,1).. (1.5,2);
     \draw[rounded corners] (0.95,1.2) .. controls(0.95,0.85).. (1.2,0.8);
     
    \draw[rounded corners] (1.2,0.6) ..controls(1.4,0.8).. (1.5,-0.7);
     \draw[rounded corners] (0.95,0.3) .. controls(1.05,0.5).. (1.2,0.6);
     
     \draw[fill=black] (0,-0.5) circle (2pt);
     \draw[fill=black] (0.2,-0.5) circle (2pt);
     \draw[fill=black] (0.95,-0.5) circle (2pt);
     \draw[fill=black] (1.48,-0.5) circle (2pt);
    \end{tikzpicture}\,,
\end{eqnarray}
where the rectangles are still to be determined, as in the step with $k=2$. Also, the rectangles that symmetrize the $k$ strands are in principle not necessarily equal to the symmetrizer obtained from the inductive step at $k$. Let us now apply the symmetrizer twice, to obtain the diagrammatic equation
\begin{eqnarray}
    \begin{tikzpicture}
    [scale=0.7,baseline={([yshift=-0.1cm]current bounding box.center)},vertex/.style={anchor=base,
    	circle,fill=black!25,minimum size=18pt,inner sep=2pt}]]
    	\draw (0,2) -- (0,1);
        \draw (0.2,2) -- (0.2,1);
        \draw[dashed] (0.25,1.5) -- (0.95,1.5);
        \draw (1,2) -- (1,1);
        \draw (-0.2,1) rectangle (1.2,0.5);
        \draw (0,0.5) -- (0,-0.5);
        \draw (0.2,0.5) -- (0.2,-0.5);
        \draw[dashed] (0.25,0) -- (0.95,0);
        \draw (1,0.5) -- (1,-0.5);
        \draw[fill=black] (0,0) circle (2pt);
        \draw[fill=black] (0.2,0) circle (2pt);
        \draw[fill=black] (1,0) circle (2pt);
        
        \draw (0,3) -- (0,4);
        \draw (0.2,3) -- (0.2,4);
        \draw[dashed] (0.25,2.5) -- (0.95,2.5);
        \draw (1,3) -- (1,4);
        \draw (-0.2,3) rectangle (1.2,2.5);
        \draw (0,2.5) -- (0,1.5);
        \draw (0.2,2.5) -- (0.2,1.5);
        \draw[dashed] (0.25,2) -- (0.95,2);
        \draw (1,2.5) -- (1,1.5);
        \draw[fill=black] (0,2) circle (2pt);
        \draw[fill=black] (0.2,2) circle (2pt);
        \draw[fill=black] (1,2) circle (2pt);
    \end{tikzpicture}
    &=&\label{eq:concatenate}
    A^2\ \ 
    \begin{tikzpicture}[scale=0.7,baseline={([yshift=-0.1cm]current bounding box.center)},vertex/.style={anchor=base,
    	circle,fill=black!25,minimum size=18pt,inner sep=2pt}]]
        \draw (0,2) -- (0,1);
        \draw (0.2,2) -- (0.2,1);
        \draw[dashed] (0.25,1.5) -- (0.9,1.5);
        \draw (0.95,2) -- (0.95,1);
        \draw (1.3,2) -- (1.3,-0.5);
        \draw (-0.2,1) rectangle (1.15,0.5);
        \draw (0,0.5) -- (0,-0.5);
        \draw (0.2,0.5) -- (0.2,-0.5);
        \draw (0.95,0.5) -- (0.95,-0.5);
        \draw[dashed] (0.25,0) -- (0.9,0);
        \draw[fill=black] (0,0) circle (2pt);
        \draw[fill=black] (0.2,0) circle (2pt);
        \draw[fill=black] (1.3,0) circle (2pt);
        \draw[fill=black] (0.95,0) circle (2pt);
        \draw (0,4.5) -- (0,3.5);
        \draw (0.2,4.5) -- (0.2,3.5);
        \draw[dashed] (0.25,4) -- (0.9,4);
        \draw (0.95,4.5) -- (0.95,3.5);
        \draw (1.3,4.5) -- (1.3,2);
        \draw (-0.2,3.5) rectangle (1.15,3);
        \draw (0,3) -- (0,2);
        \draw (0.2,3) -- (0.2,2);
        \draw (0.95,3) -- (0.95,2);
        \draw[dashed] (0.25,2.5) -- (0.9,2.5);
        \draw[fill=black] (0,2.5) circle (2pt);
        \draw[fill=black] (0.2,2.5) circle (2pt);
        \draw[fill=black] (0.95,2.5) circle (2pt);
    \end{tikzpicture}
    + 
    AB\ \
    \begin{tikzpicture}[scale=0.7,baseline={([yshift=-0.2cm]current bounding box.center)},vertex/.style={anchor=base,
    	circle,fill=black!25,minimum size=18pt,inner sep=2pt}]]
    \draw (0,2) -- (0,1.7);
    \draw (0.2,2) -- (0.2,1.7);
    \draw (0.95,2) -- (0.95,1.7);
    
    \draw[dashed] (0.3,1.9) -- (0.9,1.9);
    
    \draw (-0.2,1.7) rectangle (1.15,1.2);
    
        \draw (0,1.2) -- (0,0.3);
        \draw (0.2,1.2) -- (0.2,0.3);
        
        \draw[dashed] (0.3,0.75) -- (0.9,0.75);
    
    \draw (-0.2,0.3) rectangle (1.15,-0.2);
    
    \draw (0,-0.2) -- (0,-0.7);
    \draw (0.2,-0.2) -- (0.2,-0.7);
    \draw[dashed] (0.3,-0.45) -- (0.95,-0.45);
    \draw (0.95,-0.2) -- (0.95,-0.7);

     \draw[rounded corners] (1.2,0.8) ..controls(1.4,1).. (1.5,2);
     \draw[rounded corners] (0.95,1.2) .. controls(0.95,0.85).. (1.2,0.8);
     
    \draw[rounded corners] (1.2,0.6) ..controls(1.4,0.8).. (1.5,-0.7);
     \draw[rounded corners] (0.95,0.3) .. controls(1.05,0.5).. (1.2,0.6);
     
     \draw[fill=black] (0,-0.5) circle (2pt);
     \draw[fill=black] (0.2,-0.5) circle (2pt);
     \draw[fill=black] (0.95,-0.5) circle (2pt);
     \draw[fill=black] (1.48,-0.5) circle (2pt);
     
        \draw (0,4.5) -- (0,3.5);
        \draw (0.2,4.5) -- (0.2,3.5);
        \draw[dashed] (0.25,4) -- (0.9,4);
        \draw (0.95,4.5) -- (0.95,3.5);
        \draw (1.5,4.5) -- (1.5,2);
        \draw (-0.2,3.5) rectangle (1.15,3);
        \draw (0,3) -- (0,2);
        \draw (0.2,3) -- (0.2,2);
        \draw (0.95,3) -- (0.95,2);
        \draw[dashed] (0.25,2.5) -- (0.9,2.5);
        \draw[fill=black] (0,2.5) circle (2pt);
        \draw[fill=black] (0.2,2.5) circle (2pt);
        \draw[fill=black] (0.95,2.5) circle (2pt);
         \draw[fill=black] (1.5,2.5) circle (2pt);
    \end{tikzpicture}
    \\ \notag
    \\ \notag
    &+& 
    AB 
    \begin{tikzpicture}[scale=0.7,baseline={([yshift=-0.2cm]current bounding box.center)},vertex/.style={anchor=base,
    	circle,fill=black!25,minimum size=18pt,inner sep=2pt}]]
    \draw (0,2) -- (0,1.7);
    \draw (0.2,2) -- (0.2,1.7);
    \draw (0.95,2) -- (0.95,1.7);
    
    \draw[dashed] (0.3,1.9) -- (0.9,1.9);
    
    \draw (-0.2,1.7) rectangle (1.15,1.2);
    
        \draw (0,1.2) -- (0,0.3);
        \draw (0.2,1.2) -- (0.2,0.3);
        
        \draw[dashed] (0.3,0.75) -- (0.9,0.75);
    
    \draw (-0.2,0.3) rectangle (1.15,-0.2);
    
    \draw (0,-0.2) -- (0,-0.7);
    \draw (0.2,-0.2) -- (0.2,-0.7);
    \draw[dashed] (0.3,-0.45) -- (0.95,-0.45);
    \draw (0.95,-0.2) -- (0.95,-0.7);

     \draw[rounded corners] (1.2,0.8) ..controls(1.4,1).. (1.5,2);
     \draw[rounded corners] (0.95,1.2) .. controls(0.95,0.85).. (1.2,0.8);
     
    \draw[rounded corners] (1.2,0.6) ..controls(1.4,0.8).. (1.5,-0.7);
     \draw[rounded corners] (0.95,0.3) .. controls(1.05,0.5).. (1.2,0.6);
     
     \draw[fill=black] (0,-0.5) circle (2pt);
     \draw[fill=black] (0.2,-0.5) circle (2pt);
     \draw[fill=black] (0.95,-0.5) circle (2pt);
     \draw[fill=black] (1.48,-0.5) circle (2pt);
     \draw[fill=black] (1.45,1.5) circle (2pt);
     
        \draw (0,-0.5) -- (0,-1);
        \draw (0.2,-0.5) -- (0.2,-1);
        \draw[dashed] (0.25,-1) -- (0.9,-1);
        \draw (0.95,-0.5) -- (0.95,-1);
        \draw (1.5,-0.5) -- (1.5,-2.5);
        \draw (-0.2,-1) rectangle (1.15,-1.5);
        \draw (0,-1.5) -- (0,-2.5);
        \draw (0.2,-1.5) -- (0.2,-2.5);
        \draw (0.95,-1.5) -- (0.95,-2.5);
        \draw[dashed] (0.25,-2) -- (0.9,-2);
        \draw[fill=black] (0,-2) circle (2pt);
        \draw[fill=black] (0.2,-2) circle (2pt);
        \draw[fill=black] (0.95,-2) circle (2pt);
    \end{tikzpicture}
    + 
    B^2\ \ 
    \begin{tikzpicture}[scale=0.7,baseline={([yshift=-0.2cm]current bounding box.center)},vertex/.style={anchor=base,
    	circle,fill=black!25,minimum size=18pt,inner sep=2pt}]]
    \draw (0,2) -- (0,1.7);
    \draw (0.2,2) -- (0.2,1.7);
    \draw (0.95,2) -- (0.95,1.7);
    
    \draw[dashed] (0.3,1.9) -- (0.9,1.9);
    
    \draw (-0.2,1.7) rectangle (1.15,1.2);
    
        \draw (0,1.2) -- (0,0.3);
        \draw (0.2,1.2) -- (0.2,0.3);
        
        \draw[dashed] (0.3,0.75) -- (0.9,0.75);
    
    \draw (-0.2,0.3) rectangle (1.15,-0.2);
    
    \draw (0,-0.2) -- (0,-0.7);
    \draw (0.2,-0.2) -- (0.2,-0.7);
    \draw[dashed] (0.3,-0.45) -- (0.95,-0.45);
    \draw (0.95,-0.2) -- (0.95,-0.7);

     \draw[rounded corners] (1.2,0.8) ..controls(1.4,1).. (1.5,2);
     \draw[rounded corners] (0.95,1.2) .. controls(0.95,0.85).. (1.2,0.8);
     
    \draw[rounded corners] (1.2,0.6) ..controls(1.4,0.8).. (1.5,-0.7);
     \draw[rounded corners] (0.95,0.3) .. controls(1.05,0.5).. (1.2,0.6);
     
     \draw[fill=black] (0,-0.5) circle (2pt);
     \draw[fill=black] (0.2,-0.5) circle (2pt);
     \draw[fill=black] (0.95,-0.5) circle (2pt);
     \draw[fill=black] (1.48,-0.5) circle (2pt);
     
     
    \draw (0,4.7) -- (0,4.4);
    \draw (0.2,4.7) -- (0.2,4.4);
    \draw (0.95,4.7) -- (0.95,4.4);
    
    \draw[dashed] (0.3,4.6) -- (0.9,4.6);
    
    \draw (-0.2,4.4) rectangle (1.15,3.9);
    
        \draw (0,3.9) -- (0,3);
        \draw (0.2,3.9) -- (0.2,3);
        
        \draw[dashed] (0.3,3.45) -- (0.9,3.45);
    
    \draw (-0.2,3) rectangle (1.15,2.5);
    
    \draw (0,2.5) -- (0,2);
    \draw (0.2,2.5) -- (0.2,2);
    \draw[dashed] (0.3,2.25) -- (0.95,2.25);
    \draw (0.95,2.5) -- (0.95,2);

     \draw[rounded corners] (1.2,3.5) ..controls(1.4,3.7).. (1.5,4.7);
     \draw[rounded corners] (0.95,3.9) .. controls(0.95,3.55).. (1.2,3.5);
     
    \draw[rounded corners] (1.2,3.3) ..controls(1.4,3.5).. (1.5,2);
     \draw[rounded corners] (0.95,3) .. controls(1.05,3.2).. (1.2,3.3);
     
     \draw[fill=black] (0,2.2) circle (2pt);
     \draw[fill=black] (0.2,2.2) circle (2pt);
     \draw[fill=black] (0.95,2.2) circle (2pt);
     \draw[fill=black] (1.48,2.2) circle (2pt);
    \end{tikzpicture}\ .
\end{eqnarray}
Imposing the symmetrizer to be idempotent, it follows that the term with coefficient $A$ has idempotent boxes of degree $k$, which automatically forces this to be the symmetrizer obtained for $k$, since this is unique by induction hypothesis. It also follows that $A=1$. Then, let us indicate by $\Phi$ and $\Psi$ the diagrams whose coefficients are $A$ and $B$ in Eq.~\eqref{eq:higher_smoothing}, respectively. Then, by equating the terms where the last strand is not straight, in the idempotence condition, we have the equation $B\Psi = AB (\Phi\Psi + \Psi\Phi) + B^2 \Psi^2$, {\it i.e.} the second term of Eq.~\eqref{eq:higher_smoothing} is equal to the last three terms of Eq.~\eqref{eq:concatenate}. Using the idempotence in the known results for the Jones-Wenzl projector for degree $k$, which hold true by inductive hypothesis, it now follows that the symmetrizer at degree $k+1$ coincides with the quantum Jones-Wenzl projector, where the value of $d$ is given by the quantum dimension computed in Eq.~\eqref{eq:trace_loop}. This shows that the value of the fundamental loop with group element being the first Chebyschev polynomial, along with projecting onto the loop basis, substantially determines our effective recoupling theory to be the quantum recoupling theory of \cite{KL} at $A\neq -1$. Moreover, we observe that we have not assumed $A$ be a root of unity at any step.

We therefore obtain a diagrammatics that implies an effective recoupling, which is equivalent to the Kauffman-Lins recoupling with $A\neq -1$, namely to the recoupling theory of $SU_q(2)$ that is implemented in the Turaev-Viro model. Here strands with bullets indicate representations with the insertion of the group element, and the symmetrizer on such strands is the quantum Wenzl-Jones symmetrizer, following the procedure given above.

We emphasize that this is entirely due to the dynamical implementation of the curvature constraint at the quantum level, and the assumption that projecting onto the loop basis satisfies the idempotence condition. Thus the effective quantum representations that are found within this scheme are the by-product of the quantization of the Einstein-Hilbert action with cosmological constant in $2+1D$. \\

\section{Spin-foam dynamics} \label{sf}
 \noindent

 Quantum gravity in $2+1D$ could have been completely solved in \cite{NoPe} by regularizing the generalized projector $P$. The projection operator $P$ encodes the quantum evolution due to the presence of the Hamiltonian constraint and provides a physical scalar product in the spin-foam representation, which produces the Ponzano-Regge model. In this section we consider the procedure of extending the regularization of the projector $P$ in the presence of nontrivial cosmological constant. We start from the setting of \cite{NoPe}, and we consider the dynamics from the spin-foam perspective, as a covariant way to implement the quantization of loop quantum gravity.
 The extension of these results to the $3+1D$ case presents several difficulties that have 
 not yet been overcome \cite{Haggard:2015nat,Haggard:2014xoa,Han:2011aa,Han:2010pz,Noui:2002ag,Fairbairn:2010cp}. \\

The generalized projector $P$ defining the generic physical scalar product $<s, P\,s'>$ (between spin-network states $s$ and $s'$ of the physical Hilbert space) implements the curvature constraint in the spin-foam formalism representation of  $<s, P\,s'>$, and is expressed by

\bea \label{prolam}
&P=``\prod \limits_{x\in \Sigma} \delta\left(  \hat{F}(\omega) + \Lambda \, \hat{e} \wedge \hat{e} \right)" =\nonumber\\
& \int \mathcal{D} [N] \exp i \int_\Sigma {\rm Tr} [N \left( F(\omega)  + \Lambda \, \hat{e} \wedge \hat{e}  \right)]\,,
\eea
where $N=N^i \tau_i\in \mathfrak{su}(2)$ and $\tau_i$ are basis elements of $\mathfrak{su}(2)$ in the fundamental representation.\\

Following \cite{NoPe}, we pick a cellular decomposition $\Sigma_\delta(\Gamma, \Gamma')$ of $\Sigma$ depending on an infinitesimal parameter $\delta \in \mathbb{R}$. The cellular decomposition $\Sigma_\delta(\Gamma, \Gamma')$ consists of $0$-cells called vertices, $1$-cells that consist of edges connecting the $0$-cells, and $2$-cells called plaquettes, and denoted by $p$. The latter are squares delimited by $1$-cells between $0$-cells.
 The union of $0$-cells and $1$-cells contains the graphs $\Gamma$ and $\Gamma'$ on which the spin-networks $s$ and $s'$ are supported, respectively. Furthermore, we assume that there is a covering by open balls $\mathcal{B}_\delta$  of radii $\delta$ such that each plaquette $p$ is contained in some $\mathcal{B}_\delta$. Therefore, as $\delta \rightarrow 0$, the plaquettes shrink to points.
 This allows us to define a regularization for the physical inner product as in \cite{NoPe}. Indeed, once an  ordering for the set of plaquettes $p^i\in \Sigma_\delta(\Gamma, \Gamma')$ with $i=1,...N_p^\delta$ has been introduced ---  $N_p^\delta$ being the total number of plaquettes for an assigned value of the regulator $\delta$ of the cellular decomposition --- the physical inner product between two spin-network states $s$ and $s'$, respectively supported on $\Gamma$ and $\Gamma'$, is

\bea \label{sv}
<s\!\!\!\!&,&\!\!\! \!s'>_{\rm Phys}= <s, P s'> =\\
&=& \!\!{\rm lim}_{\delta \rightarrow 0} \sum \limits_{j_{p^i}} {\rm dim}\,j_{p^i}  <\prod_{p^i} \chi_{j_{p^i}}(U_{p^i}\, H^{-1}_\Lambda) s, s' >\,, \nonumber
\eea
where ${\rm dim}\,j_{p^i}$ stands for the dimension of the irrep of spin $j_{p^i}$, $U_{p^i}$ for the holonomy around the plaquette $p^i$, $H_\Lambda$ is a $SU(2)$ group element encoding space-time curvature and $ \chi_{j_{p^i}}(U_{p^i}\, H^{-1}_\Lambda)$ denotes the trace of the irrep $\Pi^j$ of spin $j$ of the $SU(2)$ group element ``$U_{p^i}\, H^{-1}_\Lambda$''.\\

\section{Regularization of the Hamiltonian constraint in loop basis} \label{reg}
\noindent
We can regularize the curvature constraint operator, extending the $2+1D$ analysis for the Einstein-Hilbert action developed in \cite{NoPe} to the case encoding the cosmological constant.

The main difference with respect to the previous literature is that we first smear at the classical level triads on the edge of the lattice dual to the the square tessellation of the space surfaces, and then obtain an element of the $\mathfrak{su}(2)$ algebra. We then regularize the curvature constraint as the product of two $SU(2)$ group elements, and implement the quantization in terms of holonomy operators in the fundamental representation. The effective (``quantum group'' like) recoupling theory that is induced by the quantum dynamics, as previously commented, will then extend the result to loops in any irrep of $SU(2)$. \\

In more detail, we can motivate Eq.~(\ref{sv}) by considering that, for a local patch $X\in \Sigma$ in which the cellular decomposition is made out of square cells of coordinate length $\delta$, the curvature constraint reads
\bea
\!\!\!\!F_{\Lambda}[N]&\!=\!&\int_X {\rm Tr} [N \,( F(\omega)+ \Lambda \, e \wedge e)]=\nonumber\\
&\!=\!&\!\lim \limits_{\delta \rightarrow 0}\, \sum \limits_{p^i} \delta^2\, {\rm Tr}[N_{p^i}( F_{p^i} + \Lambda\, n_{p^i} )]\,,
\eea
in which the $\mathfrak{su}(2)$ algebra elements $N_{p^i}$ and $F_{p^i}$ and $n_{p^i}$ stand respectively for the value of $N=N^j\, \tau_j$ and $\tau_j \epsilon^{ab} F_{ab}^j(\omega)$ and $\epsilon^{ab}e_a^i e_b^k \epsilon_{ik}^{\,\,\,\,\,j} \tau_j$ at an interior point $p^i$ of the $i$-th plaquette. It was already noticed in \cite{NoPe} that the holonomy $U_{p^i}$ undergoes in the $\delta\rightarrow 0$ limit the approximation 
\be
U_{p^i}(\omega)= 1\!\!1+ \delta^2 F_{p^i}(\omega)+ O(\delta^2)\,,
\ee
and that as a consequence 
\be
F[N]=\int_X {\rm Tr} [N F(\omega)]=\lim \limits_{\delta \rightarrow 0} \sum \limits_{p^i} {\rm Tr}[N_{p^i} U_{p^i}(\omega)]\,.
\ee
We notice that a similar argument can be deployed to recast the term $\Lambda \epsilon_{jk}^{i} e^j_a \,e^k_b$, which amounts to the action of the triads on the dual lattice, given that we perform the expansion of triads around the base point $p^i$ positioned at the center of the $i$-th plaquette as it follows:
\be
e^i_a(x)\big|_{p^i}\simeq \delta^i_a + O(\delta). \label{exe}
\ee
This observation allows us to define a $SU(2)$ group element $H^{\Lambda}_{p^i}$, which expanding in the infinitesimal $\delta$ parameter reads
\be \label{expH}
H^{\Lambda}_{p^i}= 1\!\!1+ \delta^2 \Lambda n_{p^i} + O(\delta^2)\,.
\ee
Within this expression the smeared (on the $\mathfrak{su}(2)$ algebra element $N$) flux of $\epsilon^{ab}e_a^i e_b^k \epsilon_{ik}^{\,\,\,\,\,j} \tau_j$ appears. Indeed 
\bea
\Lambda \int_X  {\rm Tr} [N e\wedge e]&=& \Lambda  \int_X \epsilon_{ijk} N^i \epsilon^{ab}\, e_a^j e_b^k =\nonumber\\
&=&\lim \limits_{\delta \rightarrow 0} \sum \limits_{p^i} {\rm Tr}[N_{p^i} H^\Lambda_{p^i}(\omega)]\,,
\eea
provided that on the loop states
\be \label{smd}
\delta^2 n^j_{p^i} \sim 
\lim \limits_{\delta \rightarrow 0} \Phi_{X} (\tilde{E}^j)=\lim \limits_{\delta \rightarrow 0} \int_{X}  \epsilon^{ab}e_a^i e_b^k \epsilon_{ik}^{\,\,\,\,\,j} \,.
\ee
In Eq.~\eqref{smd} we have used the definition $E^b_i=\epsilon^{ab} e^s_a \eta_{is}$ of the Ashtekar electric field on the $2$-dimensional surface of pull-back $\Sigma$, we have performed the triadic projection with respect to $e^k_b$ and finally contracted the internal indices with the Levi-Civita tensor $\epsilon_{k}^{\;\;ij}$, namely $\tilde{E}^j=E^b_i e_b^k \epsilon_{k}^{\;\;ij}$.  
Because of the gauge invariance of $\chi_{j_{p^i}}(U_{p^i}\, H^{-1}_\Lambda)$, we can rewrite the $SU(2)$ group element encoding the space-time curvature as 
\be
H^\Lambda_{p^i}=\exp (\Lambda n^3_{p^i} \tau_3)\,. 
\ee
At each plaquette we can renormalize $\Lambda$, and rewrite $H^{\Lambda'}_{p^i}=\exp (\Lambda_{p^i}' \tau_3)$. For the regularization of the projector on the loop basis we can write $H^{\Lambda'}_{p^i}=\exp (\Lambda' \tau_3)$, where the discretization is independent on the $i$-th plaquette that has been chosen. We then notice that a rescaling on the connection $\omega$ by $1/G$, so to make it dimensionless, as well as a rescaling of the coordinates by $1/\sqrt{\Lambda}$, allows to recast the action in terms of only dimensionless quantities, as  

\be\label{res_act}
S'=\frac{1}{G\, \sqrt{\Lambda}} \int_{\mathcal{M}}  \frac{1}{G\, \sqrt{\Lambda}} \, {\rm Tr}[{e} \wedge F({\omega})] + \frac{1}{3} \,{\rm Tr}[e \wedge e \wedge e]\,.
\ee
The peculiarity of Eq.~\eqref{res_act} now traces back at the level of the spin-foam dynamics to the definition of a $SU(2)$ group element encoding space-time curvature of the form $H^{\Lambda}\!=\!\exp (G \sqrt{\Lambda}\, \tau_3 n^3)$. Finally, in Planck units, i.e. with $G=1$, the curvature group element introduced in Eq.~\eqref{eq:trace_loop} is recovered, i.e.
\begin{equation}\label{eq:HLambda}
H^{\Lambda}=e^{\sqrt{\Lambda}\, \tau_3 n^3} \,.
\end{equation}
For a generic ``quantum-group effective'' irrep $j$, using the effectively induced Jones-Wenzl projector, the evaluation of the trace of $H_{\Lambda}$ provides the Chebyschev $\Delta_{2j}^{\Lambda}$ polynomial of degree-$2j$, evaluated in $\sqrt{\Lambda}$. \\

We finally comment that the curvature group element $H_\Lambda$ converges to the unity of the group $U=e$ in the vanishing cosmological constant limit $\Lambda \rightarrow 0$. Hence the standard flat curvature constraint is recovered \cite{NoPe}, which induces the convergence of the recoupling theory of $SU_q(2)$ to the standard recoupling theory of $SU(2)$.\\

\section{Two loops calculation} \label{2l}
\noindent 
As an instructive study case, we inspect the physical scalar product of two loops, for the case without cosmological constant, as in the framework of \cite{NoPe}, and subsequently repeat the calculation for the case with non-vanishing cosmological constant. 

For two cylindrical functions $\Psi_{\Gamma_1, f[A]}$ and $\Psi_{\Gamma_2, g[A]}$, the inner product is defined by the AL measure
\begin{eqnarray}
\mu_{AL}(\overline{\Psi_{\Gamma_1, f[A]}}\Psi_{\Gamma_2, g[A]})=\left\langle \Psi_{\Gamma_1, f[A]},\Psi_{\Gamma_2, g[A]}\right\rangle\\
=\int \Pi_{i=1}dh_i\overline{f(h_{\gamma_1},....,h_{\gamma_{N_{\ell}}})}g(h_{\gamma_1},....,h_{\gamma_{N_{\ell}}}),
\end{eqnarray} 
where $dh_i$ corresponds to the invariant $SU(2)$ Haar measure.

\begin{figure}[h!]
\centering
 \includegraphics[scale=0.4]{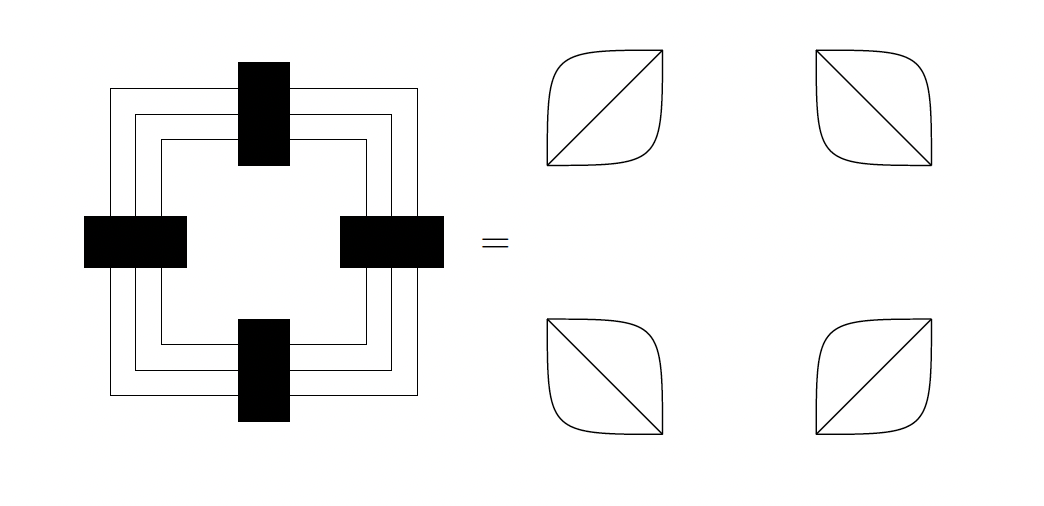}
\caption{\footnotesize{Diagrammatic definition of inner product on two loops (left) and result of integration (right). }} \label{fig:Two_loops}
\end{figure}

We calculate the inner product between two loops with spin $j$ and $j'$, inserting an extra loop corresponding to the projector of \cite{NoPe}. Specifically, we compute
\begin{eqnarray}
\left\langle O_j, O{j'}\right\rangle_{\rm Ph} &=& \int dU\Sigma_k \Delta_k\chi^*_j(U)\chi_{j'}(U)\chi_k(U)\\
\nonumber &=& \Sigma_k\Delta_k \int dU \stackrel{j}{\Pi^*}\!\!\!\!\!\!\phantom{\Pi}^{\alpha}_{\alpha}(U)\stackrel{j}{\Pi^*}\!\!\!\!\!\!\phantom{\Pi}^{\beta}_{\beta}(U)\stackrel{j'}\Pi\!\!\!\!\!\!\phantom{\Pi}^{\gamma}_{\gamma}(U) \,.\\
\end{eqnarray}


Notice that such integration can be directly solved without expanding the Dirac delta on the group, but in stead imposing the curvature constraint. In absence of cosmological constant, this latter reads $U=e$, with $U$ group element around the loop and $e$ unit element of $SU(2)$. In this case, using the composition rule $\chi^*_j(U)\chi_{j'}(U)=\sum_k \chi_k(U)$, where the sum is over the compatible spin $k$ such that $|j-j'|<k<j+j'$, we can immediately find 
\begin{eqnarray}
\left\langle O_j, O{j'}\right\rangle_{\rm Ph} &=& \int dU \chi^*_j(U)\chi_{j'}(U)\delta(U)\\
\nonumber &=& \Sigma_k \int dU \chi_k(U) \delta(U)= \Sigma_k \Delta_k \,.
\end{eqnarray}
We can now show that the same result can be recovered by expanding the Dirac delta function, as in \cite{NoPe}. In this case, integration over the three representations of the group elements along each link of the squared loop provide four pairs of trivalent intertwiners. This is shown on the right hand side of Figure~\ref{fig:Two_loops}. We represent these intertwiners in terms of Jones-Wenzl projectors and renormalize the trivalent vertices by the $\theta$-net evaluations. Specifically, the internal gauge indices of these tensors are contracted with one another according to a specific combinatorial path of contractions that respect the symmetries of the Jones-Wenzl projector, namely 
\begin{eqnarray} \label{C6}
\int &dU&  \stackrel{j}{\Pi^*}\!\!\!\!\!\!\phantom{\Pi}^{\alpha}_{\alpha}(U)\stackrel{j'}{\Pi}\!\!\!\!\!\!\phantom{\Pi}^{\beta}_{\beta}(U)\stackrel{j''}\Pi\!\!\!\!\!\!\phantom{\Pi}^{\gamma}_{\gamma}(U)
= \upsilon^{\alpha \beta \gamma}\  \upsilon_{\alpha \beta \gamma} \nonumber \\
&=& \frac{\overline{\upsilon}^{\alpha \beta \gamma}}{\sqrt{\theta(a,b,c)}} \, \  \frac{\overline{\upsilon}_{\alpha \beta \gamma}}{\sqrt{\theta(a,b,c)}} \,
\,, 
\end{eqnarray}
where $\upsilon$ is the trivalent intertwiner among the $j$, $j'$ and $j''$ representations --- assumed to be compatible to provide a non-trivial non-vanishing result, namely $j+j'+j''=\mathbb{N}$ --- of the spin-network basis, and $\overline{\upsilon}$ denotes the trivalent intertwiner in the Kauffman-Lins formalism among $a$, $b$ and $c$ fundamental representations, having introduced $a=2j$, $b=2j'$ and $c=2j''$. 

Finally, since contraction over the indices provides
\begin{eqnarray}
\overline{\upsilon}^{\alpha \beta \gamma} \overline{\upsilon}_{\alpha \beta \gamma} = \theta(a,b,c)\,, 
\end{eqnarray}
we obtain the result 
\begin{eqnarray}
\left\langle O_j, O_{j'}\right\rangle_{\rm Ph} =  \Sigma_k \Delta_k \,,
\end{eqnarray}
where the sum is over all the $k$ under the restrictions imposed by the compatibility conditions. We observe that these are finitely many, thus the result is finite.  

We now repeat the same steps, accounting for a non-vanishing cosmological constant. 

\begin{figure}[h!]  
\centering
 \includegraphics[scale=0.4]{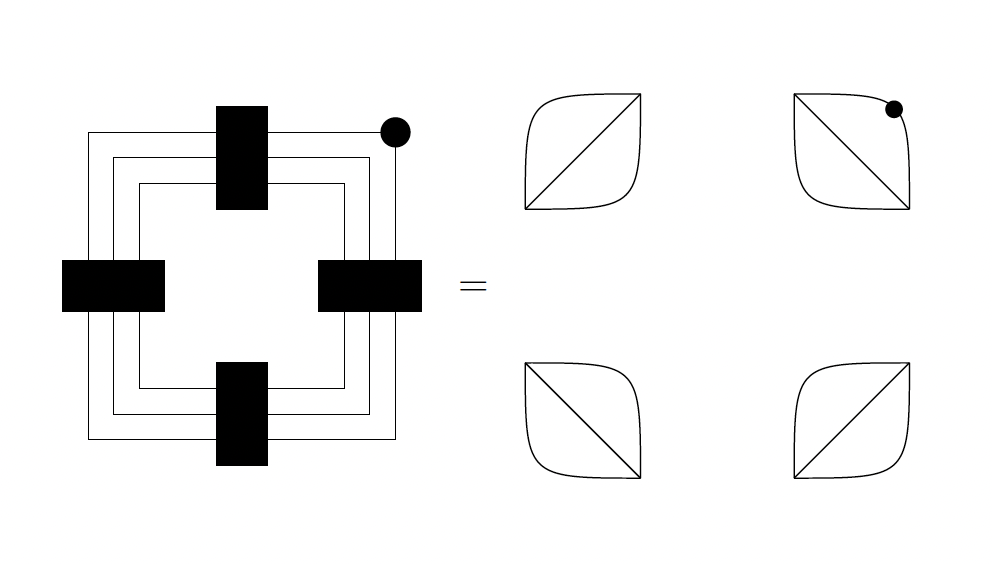} 
\caption{\footnotesize{Diagrammatic definition of inner product on two loops (left) and result of integration (right), where the black dot indicates the presence of the group element.}}
\label{fig:Lambda_two_loops}
\end{figure}

We calculate the inner product between two loops with spin $j$ and $j'$
\begin{eqnarray}
\left\langle O_j, O_{j'}\right\rangle_{\rm Ph} &=& \int dU\Sigma_k \Delta_k\chi^*_j(U)\chi_{j'}(U)\chi_k(UH_{\Lambda}^{-1})\\
\nonumber &=& \Sigma_k\Delta_k \int dU \stackrel{j}{\Pi^*}\!\!\!\!\!\!\phantom{\Pi}^{\alpha_1}_{\beta_1}(U)\stackrel{j'}{\Pi^*}\!\!\!\!\!\!\phantom{\Pi}^{\beta_1}_{\alpha_1}(U)\stackrel{j'}\Pi\!\!\!\!\!\!\phantom{\Pi}^{\alpha_2}_{\beta_2}(U)\\
&\phantom{a}&  \qquad \qquad  \ \ \times \stackrel{j'}{\Pi}\!\!\!\!\!\!\phantom{\Pi}^{\beta_2}_{\alpha_2}(U) \stackrel{k}{\Pi}\!\!\!\!\!\!\phantom{\Pi}^{\alpha_3}_{\beta_3}(U)\stackrel{k}{\Pi}\!\!\!\!\!\!\phantom{\Pi}^{\beta_3}_{\alpha_3}(H^{-1}_{\Lambda}), \nonumber \\
\notag &=& \Sigma_k\Delta_k \int dU \stackrel{j}{\Pi^*}\!\!\!\!\!\!\phantom{\Pi}^{\alpha_1}_{\beta_1}(U)\stackrel{j'}{\Pi}\!\!\!\!\!\!\phantom{\Pi}^{\beta_2}_{\alpha_2}(U)\stackrel{k}{\Pi}\!\!\!\!\!\!\phantom{\Pi}^{\alpha_3}_{\beta_3}(U) \nonumber \\
&\phantom{a}&  \qquad \qquad  \ \ \times
\stackrel{k}{\Pi}\!\!\!\!\!\!\phantom{\Pi}^{\beta_3}_{\alpha_3}(H^{-1}_{\Lambda})\frac{\delta_{jj'}}{\Delta_j}\delta_{\beta_1\beta_2}\delta_{\alpha_1\alpha_2}\,, \nonumber
\end{eqnarray}
where we have used
\begin{eqnarray}
\stackrel{j}{\Pi^*}\!\!\!\!\!\!\phantom{\Pi}^{\alpha}_{\beta}(U)\stackrel{j'}{\Pi}\!\!\!\!\!\!\phantom{\Pi}^{\gamma}_{\delta}(U)=\frac{\delta_{jj'}}{\Delta_j}\delta_{\alpha\delta}\delta_{\beta\gamma}=\delta_{jj'}=1,\label{eqn:produreps}
\end{eqnarray}
and
\begin{eqnarray}
\overset{j_p}{\chi}(U_{p}H_{\Lambda}^{-1})=\overset{j_p}\Pi\!\!\!\!\phantom{\Pi}^{\alpha}_{\beta}(U_p)\overset{j_p}{\Pi}\!\!\!\!\phantom{\Pi}^{\beta}_{\alpha}(H_{\Lambda}^{-1}).
\end{eqnarray}
The definition is represented diagrammatically in the right hand side of Figure~\ref{fig:Lambda_two_loops}.


Among the four pairs of trivalent intertwiners, only one of them will encapsulate the representations of the group element implementing the curvature constraint, namely 
\begin{eqnarray}
&&\upsilon^{\alpha_1 \alpha_2 \alpha_3}\  \overset{j''}{\Pi}\!\!\!\!\phantom{\Pi}^{\beta_3}_{\alpha_3}(H^{-1}_{\Lambda}) \upsilon_{\alpha_1 \alpha_2 \beta_3}   \\
&&=\overline{\upsilon}^{\alpha_1 \alpha_2 \alpha_3}\  \overset{\frac{c}{2}}{\Pi}\!\!\!\!\phantom{\Pi}^{\beta_3}_{\alpha_3}(H^{-1}_{\Lambda}) \  \overline{\upsilon}_{\alpha_1 \alpha_2 \beta_3} \frac{1}{ \theta(a,b,c)} 
= \frac{\theta_{\Lambda}(a,b,c)}{ \theta(a,b,c)}\,. \nonumber 
\end{eqnarray}
Making use of Lemma~7 in \cite{KL}, by inserting the group element corresponding to the cosmological constant, we obtain the expression in Figure~5, from which it is immediate to compute the value of the $\theta$-nets with one insertion of group element with cosmological constant, $H_\Lambda$. 
\begin{figure}[h!]
\centering
 \includegraphics[scale=0.3]{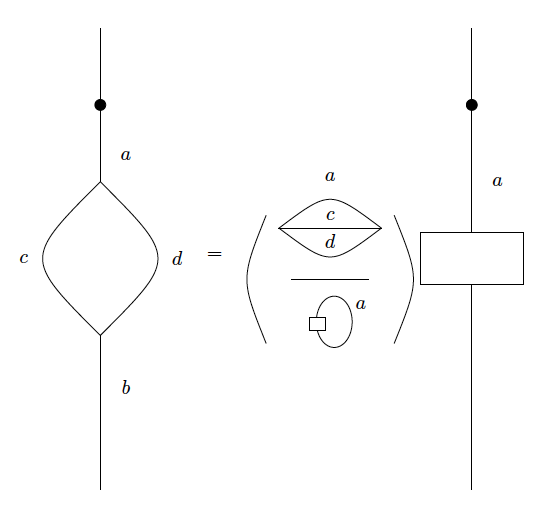}
\caption{\footnotesize{Diagrammatic reformulation of Lemma~7 in \cite{KL}, with quantum group representation here induced by the cosmological constant.}} \label{fig:Lemma7}
\end{figure}
Therefore it follows that 
\begin{eqnarray} \label{C8}
\frac{\theta_{\Lambda}(a,b,c)}{ \theta(a,b,c)} 
=\frac{\overset{k}{\chi}(H_{\Lambda}^{-1})}{\Delta_k}\,.
\end{eqnarray}
Notice that contractions inside and outside an integral follows respectively from 
\begin{eqnarray}
\int dg\overset{j}{\Pi^{\alpha_1} _{\alpha_2}}(g^{-1})\overset{j'}{\Pi^{\beta_1} _{\beta_2}}(g)=\frac{\delta_{jj'}}{\Delta j}\delta^{\alpha_1\beta_1}\delta_{\alpha_2\beta_2}\,,
\end{eqnarray}
which diagrammatically recasts into
\begin{eqnarray}
\begin{tikzpicture}[scale=0.50,baseline={([yshift=-0.1cm]current bounding box.center)},vertex/.style={anchor=base,
    	circle,fill=black!25,minimum size=18pt,inner sep=2pt}]
	    \draw (1,0.3)--(4,0.3);
	    \draw (1,0)-- (4,0);
	    
	    \filldraw [fill = black] (2.3,-0.3) rectangle (2.7,0.6);
	\end{tikzpicture} = \frac{\delta_{jj'}}{\Delta_j} \ \  \begin{tikzpicture}[scale=0.50,baseline={([yshift=-0.1cm]current bounding box.center)},vertex/.style={anchor=base,
    	circle,fill=black!25,minimum size=18pt,inner sep=2pt}]
	\draw (1,2.5) arc (270:450:0.3cm and 0.25cm);
	\draw (2,3) arc (90:270:0.3cm and 0.25cm);
	
	\draw (1,2.5) -- (0.5,2.5);
	\draw (1,3) -- (0.5,3);
	
	\draw (2,3) -- (2.5,3);
	\draw (2,2.5) -- (2.5,2.5);
	\end{tikzpicture}
	\,,
\end{eqnarray}
and follows from Eq.~(\ref{eqn:produreps}).
\\

Finally, taking into account Eqs. \eqref{C6} and \eqref{C8}, we are able to recover the evaluation of the inner product, as in
\begin{eqnarray}
\left\langle O_j, O_{j'}\right\rangle &=& \int dU\Sigma_k \Delta_k\chi^*_j(U)\chi_{j'}(U)\chi_k(UH_{\Lambda}^{-1})\nonumber \\
&=&  \sum_k \Delta_k \frac{\overline{\upsilon} \cdot \overline{\upsilon}}{\theta(a,\,b,\,c)} \, \frac{{\chi_k}(H_{\Lambda})}{\Delta_k} \nonumber \\
&=&  \sum_k  {\chi_k}(H_{\Lambda}) \,.
\end{eqnarray}
We observe that this is formally the same as in the case without cosmological constant, where the classical dimension has been replaced by the quantum dimension, as calculated in Eq.~(\ref{eq:trace_loop}).  Furthermore, we observe that this procedure, by expanding the Dirac delta function in representations of $SU(2)$, retains a spurious dependence on the $SU(2)$ group elements that might eventually render more difficult the interpretation of the results in term of the recoupling theory of $SU_q(2)$.\\

A more intuitive path to recognize the emergence of  the recoupling theory of $SU_q(2)$ amounts to directly integrating out the $SU(2)$ elements. This corresponds, from a physical perspective, to imposing the constraints at the quantum level on the loop elements.

\begin{center}
\begin{figure}[h!]
   \begin{tikzpicture}[scale=1.3]
\draw (0,0) circle (20pt);
\draw (0,0) circle (25pt);

\draw[fill=black] (-0.87,0) circle (2pt);
\draw[fill=black] (-0.70,0) circle (2pt);
\end{tikzpicture} 
\caption{Two loops with only group element $H_\Lambda$ inserted, as resulting from applying the Dirac delta-function imposing the curvature constraint.}
\label{fig:loops_with_lambda}
\end{figure}
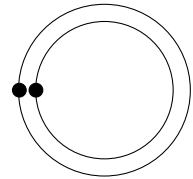
\end{center}

As in the standard case, we may opt for imposing the curvature constraint without expanding the Dirac delta function. In this case:
\begin{eqnarray}
\left\langle O_j, O_{j'}\right\rangle &=& \int dU \chi^*_j(U)\chi_{j'}(U)\delta(UH_{\Lambda})\nonumber \\
&=&  \sum_k \int dU \chi_k(U)
\delta(U H_{\Lambda}^{-1})
\nonumber \\
&=&  \sum_k  \chi_k(H_{\Lambda}) \,,
\end{eqnarray}
see Figure~\ref{fig:loops_with_lambda}.

\section{Diffeomorphism invariance} \label{diffeo}
\noindent 
We can now check how the projector extended so to include the cosmological constant, namely Eq.~\eqref{prolam}, naturally incorporates diffeomorphism invariance, as for the physical projector introduced in \cite{NoPe}.

\subsection{Case without vertices involved}
\noindent 
We shall first inspect the case in which spin-network states, or their sub-states, enclose no intertwiners. In this case, considering two holonomies with different shapes, and using the spurious notation that arises from expanding the Dirac delta-functions imposing the curvature constraint, we can easily convince ourselves that

\begin{eqnarray}
&\sum_k \Delta_k& 
\begin{tikzpicture}[scale=0.50,baseline={([yshift=0cm]current bounding box.center)},vertex/.style={anchor=base,
    	circle,fill=black!25,minimum size=18pt,inner sep=2pt}]
		\draw (0,3) -- (7,3);
		\draw (0,2.5) -- (2,2.5);
		\draw (2,2.5) arc (180:360:1.5cm and  1.5cm);
		\draw (5,2.5) -- (7,2.5);
		
		\draw (2.5,2.5) arc (180:360:1cm  and 1.2cm);
		
		\draw (2.5,2.5) -- (4.5,2.5);
		
		\filldraw [fill = black] (1,2.2) rectangle (1.3,3.2);
		\filldraw [fill = black] (3.4,0.7) rectangle (3.7,1.5);
		\filldraw [fill = black] (3.4,2.2) rectangle (3.7,3.2);
		\filldraw [fill = black] (5.8,2.2) rectangle (6.1,3.2);
		\filldraw[fill=black] (4.5,2.5) circle (3pt) ;
		
		\node (a)  at (0.5,3.6) {$j$};
		\node (a)  at (6.5,3.6) {$j$};
		\node (a)  at (3,2) {$k$};
		\end{tikzpicture}\\
&=& \sum_k \frac{\Delta_k}{\Delta_j^3}\ \begin{tikzpicture}[scale=0.50,baseline={([yshift=0cm]current bounding box.center)},vertex/.style={anchor=base,
    	circle,fill=black!25,minimum size=18pt,inner sep=2pt}]
	\draw (1,2.5) arc (270:450:0.3cm and 0.25cm);
	\draw (6,3) arc (90:270:0.3cm and 0.25cm);
	
	\draw (1,2.5) -- (0.5,2.5);
	\draw (1,3) -- (0.5,3);
	
	\draw (6,3) -- (6.5,3);
	\draw (6,2.5) -- (6.5,2.5);
	
	\draw (2,2.5) arc (180:360:1.5cm and  1.5cm);
	\draw (2.5,2.5) arc (180:360:1cm  and 1.2cm);
	
	\draw (2,3) arc (90:270:0.3cm and 0.25cm);
	\draw (5,2.5) arc (270:450:0.3cm and 0.25cm);
	
	\draw (2,3) -- (3,3);
	\draw (2.5,2.5) --(3,2.5);
	
	\draw (3,2.5) arc (270:450:0.3cm and 0.25cm);

	\draw (4.5,2.5) -- (4,2.5);
	\draw (5,3) -- (4,3);
	\draw (4,3) arc (90:270:0.3cm and 0.25cm);
	
	\filldraw[fill=black] (4.5,2.5) circle (3pt);
	
		\filldraw [fill = black] (3.4,0.7) rectangle (3.7,1.5);
		
		\node (a)  at (0.5,3.6) {$j$};
		\node (a)  at (6.5,3.6) {$j$};
		\node (a)  at (3,2) {$k$};
	\end{tikzpicture} \\
&=& \sum_k \delta_{jk}\frac{\Delta_k}{\Delta^4_j}\ \begin{tikzpicture}[scale=0.50,baseline={([yshift=-0.1cm]current bounding box.center)},vertex/.style={anchor=base,
    	circle,fill=black!25,minimum size=18pt,inner sep=2pt}]
	\draw (1,2.5) arc (270:450:0.3cm and 0.25cm);
	\draw (6,3) arc (90:270:0.3cm and 0.25cm);
	
	\draw (1,2.5) -- (0.5,2.5);
	\draw (1,3) -- (0.5,3);
	
	\draw (6,3) -- (6.5,3);
	\draw (6,2.5) -- (6.5,2.5);
	
	\draw (3,2.7) circle (10pt);
	\draw (4.5,2.7) circle (10pt);
	
	\filldraw[fill=black] (4.2,2.5) circle (3pt);
	
	\node (a)  at (0.5,3.6) {$j$};
		\node (a)  at (6.5,3.6) {$j$};
		\node (a)  at (4.5,2) {$k$};
		\node (a)  at (3.5,2) {$j$};
	\end{tikzpicture}\\
	&=& \frac{\Delta^{\Lambda}_j}{\Delta^2_j}\
    \begin{tikzpicture}[scale=0.50,baseline={([yshift=-0.1cm]current bounding box.center)},vertex/.style={anchor=base,
    	circle,fill=black!25,minimum size=18pt,inner sep=2pt}]
	\draw (1,2.5) arc (270:450:0.3cm and 0.25cm);
	\draw (2,3) arc (90:270:0.3cm and 0.25cm);
	
	\draw (1,2.5) -- (0.5,2.5);
	\draw (1,3) -- (0.5,3);
	
	\draw (2,3) -- (2.5,3);
	\draw (2,2.5) -- (2.5,2.5);
	\end{tikzpicture}\\
	&=& \frac{\Delta^{\Lambda}_j}{\Delta_j}\ \ \begin{tikzpicture}[scale=0.50,baseline={([yshift=-0.1cm]current bounding box.center)},vertex/.style={anchor=base,
    	circle,fill=black!25,minimum size=18pt,inner sep=2pt}]
	    \draw (1,0.3)--(4,0.3);
	    \draw (1,0)-- (4,0);
	    
	    \filldraw [fill = black] (2.3,-0.3) rectangle (2.7,0.6);
	\end{tikzpicture}\\
	&=& \begin{tikzpicture}[scale=0.50,baseline={([yshift=-0.1cm]current bounding box.center)},vertex/.style={anchor=base,
    	circle,fill=black!25,minimum size=18pt,inner sep=2pt}]
	    \draw (1,0.3)--(6,0.3);
	    \draw (1,0)-- (6,0);
	    
	    \filldraw[fill = black] (2.3,-0.3) rectangle (2.7,0.6);
	    \filldraw[fill=black] (3.5,0) circle (2pt);
	    \filldraw[fill = black] (4.3,-0.3) rectangle (4.7,0.6);
	\end{tikzpicture}
	 \,.
\end{eqnarray}

\subsection{Case with vertices involved}
\noindent 
We now consider the case of diffeomorphism invariance when spin-network states contain vertices. First, we observe that due to the translation invariance of the Haar measure, it follows that the following (diagrammatic) equations hold
\begin{eqnarray}\label{eqn:box_slide}
   \begin{tikzpicture}[scale=0.3,baseline={([yshift=-0.1cm]current bounding box.center)},vertex/.style={anchor=base,
    	circle,fill=black!25,minimum size=18pt,inner sep=2pt}]
    \draw (0,1.5) -- (5,1.5);
    \draw (0,0) -- (5,0);
    \draw[fill=black] (2.2,-0.5) rectangle (2.8,2);
    \filldraw[fill=black] (1,0) circle (5pt);
\end{tikzpicture}\  \ &=&
\begin{tikzpicture}[scale=0.3,baseline={([yshift=-0.1cm]current bounding box.center)},vertex/.style={anchor=base,
    	circle,fill=black!25,minimum size=18pt,inner sep=2pt}]
    \draw (0,1.5) -- (5,1.5);
    \draw (0,0) -- (5,0);
    \draw[fill=black] (2.2,-0.5) rectangle (2.8,2);
    \filldraw[fill=black] (4,1.5) circle (5pt);
\end{tikzpicture}
\end{eqnarray}

\begin{eqnarray}\label{eqn:vertex_slide}
\begin{tikzpicture}[scale=0.3,baseline={([yshift=-0.1cm]current bounding box.center)},vertex/.style={anchor=base,
    	circle,fill=black!25,minimum size=18pt,inner sep=2pt}]
        \draw (-3,1.5) -- (0,0) -- (3,1.5);
        \draw (0,0) -- (0,-3);
        \filldraw[fill=black] (0,-2) circle (5pt);
\end{tikzpicture}\ \ &=& 
\begin{tikzpicture}[scale=0.3,baseline={([yshift=-0.1cm]current bounding box.center)},vertex/.style={anchor=base,
    	circle,fill=black!25,minimum size=18pt,inner sep=2pt}]
        \draw (-3,1.5) -- (0,0) -- (3,1.5);
        \draw (0,0) -- (0,-3);
        \filldraw[fill=black] (-1,0.5) circle (5pt);
        \filldraw[fill=black] (1,0.5) circle (5pt);
\end{tikzpicture}
\end{eqnarray}
where the group element on the left hand side of Eq.~\eqref{eqn:vertex_slide} gives two copies of its inverse on the right hand side. Consider a transition of type
\begin{center}
\begin{tikzpicture}[scale=0.4]
     \draw (0,0) -- (7,0);
     \draw (0,0.5) -- (4,0.5);
     \draw (4,0.5) -- (4.2,0.3);
     \draw (4.5,-0.2) -- (8,-4);
     \draw (4,0.5) -- (8,4.5);
     \draw (7,-2) -- (7,2.5);
     \draw (7,2.5) -- (8.5,4);
     \draw (7,-2) -- (8.5,-3.5);
     
     \draw[dashed] (-1.5,0.25) -- (9,0.25);
     
     \draw[fill=black] (1,-0.3) rectangle (1.5,0.8);
     \draw[fill=black,rotate=45] (7.2,-3.4) rectangle (7.7,-2.2);
     \draw[fill=black,rotate=-45] (7,3.8) rectangle (7.5,2.6);
     
     \node (a) at (2.5,1.2) {$j$};
     \node (a) at (2.5,-0.7) {$j$};
     \node (a) at (5,2.5) {$k$};
     \node (a) at (5,-2) {$k$};
     \node (a) at (7.5,1.2) {$m$};
     \node (a) at (7.5,-0.7) {$n$};
\end{tikzpicture}
\end{center}
Now, by inserting the projector with group element, we compute
\begin{eqnarray}
\lefteqn{\sum_{p,q} \Delta_p\Delta_q\ 
\begin{tikzpicture}[scale=0.4,baseline={([yshift=-0.1cm]current bounding box.center)},vertex/.style={anchor=base,
    	circle,fill=black!25,minimum size=18pt,inner sep=2pt}]
\draw (0,0.5) -- (1,0.5);
\draw (0,0) -- (1,0);
\draw[fill=black] (1,-0.2) rectangle (1.2,0.7);
\draw (1,0.5) -- (3,0.5);
\draw (1,0) -- (7.5,0);
\draw (3,0.5) -- (3.8,0.5);
\draw (4.2,-0.1) -- (8.3,-4);
\draw (3.8,0.5) -- (4.05,0.22);
\draw  (5,-0.3) -- (7,-0.3) -- (7,-2.2) -- (5,-0.3);
\draw (4.8, 0.5) -- (7,0.5) -- (7,2.3) -- (4.8,0.5);
\draw (3.8,0.5) -- (8.8,4.5);
\draw (7.5,-2.6) -- (7.5,2.75); 
\draw (7.5,2.75) -- (9.2,4.1);
\draw (7.5,-2.6) -- (8.5, -3.55);
\draw (8.5,-3.55) -- (9,-4);
\draw (8.3,-4) -- (8.7,-4.4);
\draw[fill=black] (6,-0.5) rectangle (6.3,0.7);
\draw[fill=black] (6.8,1.2) rectangle (7.7,1.5);
\draw[fill=black] (6.8,-1.2) rectangle (7.7,-1.5);
\draw[rotate=45,fill=black] (5,-2.45) rectangle (5.35,-3.35);
\draw[rotate=135,fill=black] (-5.2,-2.7) rectangle (-5.5,-3.6);
\draw[rotate=45,fill=black] (8,-2.8) rectangle (8.4,-3.65);
\draw[rotate=135,fill=black] (-8,-2.8) rectangle (-8.4,-3.65);
\node (a)  at (0.5,1) {$j$};
\node (a)  at (0.5,-0.5) {$j$};
\node (a)  at (6,2.7) {$k$};
\node (a)  at (6.5,-2.9) {$\ell$};
\node (a)  at (8,2) {$m$};
\node (a)  at (8,-2) {$n$};
\node (a)  at (6.5,1.5) {$p$};
\node (a)  at (6.5,-1) {$q$};
\draw[fill=black] (7,2.3) circle (2.5pt);
\draw[fill=black] (7,-0.3) circle (2.5pt);
\end{tikzpicture}}\label{eqn:initial_vertex} \\
&=& 
\sum_{p,q} \Delta_p\Delta_q\ \begin{tikzpicture}[scale=0.4,baseline={([yshift=-0.1cm]current bounding box.center)},vertex/.style={anchor=base,
    	circle,fill=black!25,minimum size=18pt,inner sep=2pt}]
\draw (0,0.5) -- (1,0.5);
\draw (0,0) -- (1,0);
\draw[fill=black] (1,-0.2) rectangle (1.2,0.7);
\draw (1,0.5) -- (3,0.5);
\draw (1,0) -- (7.5,0);
\draw (3,0.5) -- (3.8,0.5);
\draw (4.2,-0.1) -- (8.3,-4);
\draw (3.8,0.5) -- (4.05,0.22);
\draw  (5,-0.3) -- (7,-0.3) -- (7,-2.2) -- (5,-0.3);
\draw (4.8, 0.5) -- (7,0.5) -- (7,2.3) -- (4.8,0.5);
\draw (3.8,0.5) -- (8.8,4.5);
\draw (7.5,-2.6) -- (7.5,2.75); 
\draw (7.5,2.75) -- (9.2,4.1);
\draw (7.5,-2.6) -- (8.5, -3.55);
\draw (8.5,-3.55) -- (9,-4);
\draw (8.3,-4) -- (8.7,-4.4);
%
%
\draw[rotate=45,fill=black] (5,-2.45) rectangle (5.35,-3.35);
\draw[rotate=135,fill=black] (-5.2,-2.7) rectangle (-5.5,-3.6);
\draw[rotate=45,fill=black] (8,-2.8) rectangle (8.4,-3.65);
\draw[rotate=135,fill=black] (-8,-2.8) rectangle (-8.4,-3.65);
\node (a)  at (0.5,1) {$j$};
\node (a)  at (0.5,-0.5) {$j$};
\node (a)  at (6,2.7) {$k$};
\node (a)  at (6.5,-2.9) {$\ell$};
\node (a)  at (8,2) {$m$};
\node (a)  at (8,-2) {$n$};
\node (a)  at (6.5,1.5) {$p$};
\node (a)  at (6.5,-1) {$q$};
\draw[fill=black] (7,2.3) circle (2.5pt);
\draw[fill=black] (7,-0.3) circle (2.5pt);
\end{tikzpicture}\label{eqn:vertex_gauge_fixing} \\
&=&
\begin{tikzpicture}[scale=0.4,baseline={([yshift=-0.1cm]current bounding box.center)},vertex/.style={anchor=base,
	circle,fill=black!25,minimum size=18pt,inner sep=2pt}]
\draw (0,0.5) -- (1,0.5);
\draw (0,0) -- (1,0);
\draw[fill=black] (1,-0.2) rectangle (1.2,0.7);
\draw (1,0.5) -- (3,0.5);
\draw (1,0) -- (7.5,0);
\draw (3,0.5) -- (3.8,0.5);
\draw (4.2,-0.1) -- (5.5,-1.5);
\draw (6,-2) -- (8.3,-4);
\draw (3.8,0.5) -- (4.05,0.22);
\draw  (5,-0.3) -- (7,-0.3)--(7,-2.2);
\draw (5,-0.3)-- (5.8,-1.2);
\draw (7,-2.2) -- (6.5,-1.8);
\draw (6.5,-1.8) arc (45:210:0.3cm and 0.2cm);
\draw (5.5,-1.5) arc (225:415:0.24cm and 0.2cm);
\draw (4.8, 0.5) -- (7,0.5) -- (7,2.3);
\draw  (4.8,0.5)-- (5.7,1.2);
\draw (7,2.3) -- (6.3,1.75);
\draw (3.8,0.5) -- (5.3,1.5);
 \draw (6,2.1)--(9,4.5);
 \draw (6,2.1) arc (135:297:0.28cm and 0.21cm);
 \draw (5.7,1.2) arc (-45:150:0.25cm and 0.25cm);
\draw (7.5,-2.6) -- (7.5,2.75); 
\draw (7.5,2.75) -- (9.2,4.1);
\draw (7.5,-2.6) -- (8.5, -3.55);
\draw (8.5,-3.55) -- (9,-4);
\draw (8.3,-4) -- (8.7,-4.4);
%
%
%
\draw[rotate=45,fill=black] (8,-2.8) rectangle (8.4,-3.65);
\draw[rotate=135,fill=black] (-8,-2.8) rectangle (-8.4,-3.65);
\node (a)  at (0.5,1) {$j$};
\node (a)  at (0.5,-0.5) {$j$};
\node (a)  at (6,2.7) {$k$};
\node (a)  at (6.5,-2.9) {$\ell$};
\node (a)  at (8,2) {$m$};
\node (a)  at (8,-2) {$n$};
\node (a)  at (6.5,1.5) {$p$};
\node (a)  at (6.5,-1) {$q$};
\draw[fill=black] (7,2.3) circle (2.5pt);
\draw[fill=black] (7,-0.3) circle (2.5pt);
\end{tikzpicture}
\end{eqnarray}
\begin{eqnarray}
&=&
\begin{tikzpicture}[scale=0.4,baseline={([yshift=-0.1cm]current bounding box.center)},vertex/.style={anchor=base,
    	circle,fill=black!25,minimum size=18pt,inner sep=2pt}]
\draw (0,0.5) -- (1,0.5);
\draw (0,0) -- (1,0);
\draw[fill=black] (1,-0.2) rectangle (1.2,0.7);
\draw (1,0.5) -- (3,0.5);
\draw (1,0) -- (7.5,0);
\draw (3,0.5) -- (3.8,0.5);
\draw (4.2,-0.1) -- (8.3,-4);
%
%
\draw (3.8,0.5) -- (8.8,4.5);
\draw (3.8,0.5) -- (4.05,0.22);
\draw (7.5,-2.6) -- (7.5,2.75); 
\draw (7.5,2.75) -- (9.2,4.1);
\draw (7.5,-2.6) -- (8.5, -3.55);
\draw (8.5,-3.55) -- (9,-4);
\draw (8.3,-4) -- (8.7,-4.4);
%
%
\draw[rotate=45,fill=black] (8,-2.8) rectangle (8.4,-3.65);
\draw[rotate=135,fill=black] (-8,-2.8) rectangle (-8.4,-3.65);
\node (a)  at (0.5,1) {$j$};
\node (a)  at (0.5,-0.5) {$j$};
\node (a)  at (6,2.9) {$k$};
\node (a)  at (6.5,-2.9) {$\ell$};
\node (a)  at (8,2) {$m$};
\node (a)  at (8,-2) {$n$};
\draw[fill=black] (5,1.45) circle (2.5pt);
\draw[fill=black] (5,-0.85) circle (2.5pt);
\end{tikzpicture}\label{eqn:vertex_integration}
\\
&=&\begin{tikzpicture}[scale=0.4,baseline={([yshift=-0.1cm]current bounding box.center)},vertex/.style={anchor=base,
    	circle,fill=black!25,minimum size=18pt,inner sep=2pt}]
\draw (0,0.5) -- (1,0.5);
\draw (0,0) -- (1,0);
\draw[fill=black] (1,-0.2) rectangle (1.2,0.7);
\draw (1,0.5) -- (3,0.5);
\draw (1,0) -- (7.5,0);
\draw (3,0.5) -- (3.8,0.5);
\draw (4.2,-0.1) -- (8.3,-4);
%
%
\draw (3.8,0.5) -- (8.8,4.5);
\draw (3.8,0.5) -- (4.05,0.22);
\draw (7.5,-2.6) -- (7.5,2.75); 
\draw (7.5,2.75) -- (9.2,4.1);
\draw (7.5,-2.6) -- (8.5, -3.55);
\draw (8.5,-3.55) -- (9,-4);
\draw (8.3,-4) -- (8.7,-4.4);
%
%
\draw[rotate=45,fill=black] (8,-2.8) rectangle (8.4,-3.65);
\draw[rotate=135,fill=black] (-8,-2.8) rectangle (-8.4,-3.65);
\node (a)  at (0.5,1) {$j$};
\node (a)  at (0.5,-0.5) {$j$};
\node (a)  at (6,2.9) {$k$};
\node (a)  at (6.5,-2.9) {$\ell$};
\node (a)  at (8,2) {$m$};
\node (a)  at (8,-2) {$n$};
\draw[fill=black] (2.5,0.5) circle (2.5pt);
\end{tikzpicture}
\end{eqnarray}
\begin{eqnarray}
&=& 
\begin{tikzpicture}[scale=0.4,baseline={([yshift=-0.1cm]current bounding box.center)},vertex/.style={anchor=base,
    	circle,fill=black!25,minimum size=18pt,inner sep=2pt}]
\draw (0,0.5) -- (1,0.5);
\draw (0,0) -- (1,0);
\draw[fill=black] (1,-0.2) rectangle (1.2,0.7);
\draw (1,0.5) -- (3,0.5);
\draw (1,0) -- (7.5,0);
\draw (3,0.5) -- (3.8,0.5);
\draw (4.2,-0.1) -- (8.3,-4);
%
%
\draw (3.8,0.5) -- (8.8,4.5);
\draw (3.8,0.5) -- (4.05,0.22);
\draw (7.5,-2.6) -- (7.5,2.75); 
\draw (7.5,2.75) -- (9.2,4.1);
\draw (7.5,-2.6) -- (8.5, -3.55);
\draw (8.5,-3.55) -- (9,-4);
\draw (8.3,-4) -- (8.7,-4.4);
%
%
\draw[rotate=45,fill=black] (8,-2.8) rectangle (8.4,-3.65);
\draw[rotate=135,fill=black] (-8,-2.8) rectangle (-8.4,-3.65);
\node (a)  at (0.5,1) {$j$};
\node (a)  at (0.5,-0.5) {$j$};
\node (a)  at (6,2.9) {$k$};
\node (a)  at (6.5,-2.9) {$\ell$};
\node (a)  at (8,2) {$m$};
\node (a)  at (8,-2) {$n$};
\draw[fill=black] (1.8,0.5) circle (2.5pt);
\draw[fill=black] (2.5,-0.2) rectangle (2.7,0.7);
    %
    %
    %
    
\end{tikzpicture}
\end{eqnarray}
where we have used the gauge fixing identity (see Appendix of \cite{NoPe}) in the first equality, and then we have integrated to obtain the second equality. We observe that, upon using Eq.~\eqref{eqn:box_slide}, {\it i.e.} sliding the group elements appropriately, we can indeed obtain a configuration where we can draw loops intersecting the spin-network only through Haar integration boxes, where the group elements do not appear inside the loops; this ensures that we can apply the gauge fixing identity even though the group elements appear in the spin-network. Finally, the double integration on the left part of the last diagrammatic equality provides a factor $\Delta_j^\Lambda/\Delta_j$ that multiplies the result without cosmological constant derived in \cite{NoPe}.  \\

\section{Relation to the Turaev-Viro model} \label{tuvi}
\noindent 
We now consider the relation between the present theory and the Turaev-Viro model. In particular, we show that the quantum group recoupling theory of Kauffman and Lins \cite{KL} arises by introducing the cosmological constant in the original computation of Noui and Perez in \cite{NoPe}. Explicitly, we find that inserting the group element that arises from the cosmological constant, the tetra-net corresponding to certain transition amplitudes becomes the quantum $6j$ symbol. 

If we proceed by imposing the curvature constraint through the Dirac delta function $\delta(UH_\Lambda^{-1})$, the result is immediate:

\begin{figure}[h!]
\centering
 \includegraphics[scale=0.4]{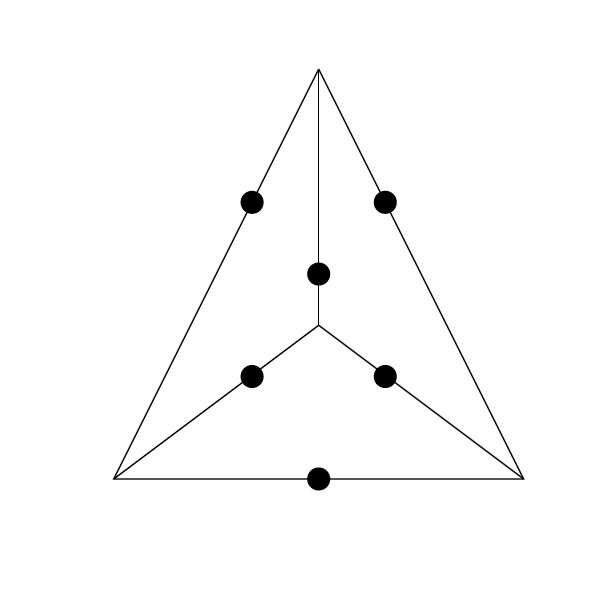}
\caption{\footnotesize{Tetra-net where non-vanishing curvature has been imposed homogeneously around the circles.}} \label{fig:Hom_tetra}
\end{figure}

where the intertwiners are now compatible with the recoupling theory of $SU_q(2)$, by assumption that the symmetrizer of irreps is a projector.

In order to evaluate this tetra-net, we employ the chromatic evaluation of \cite{KL}, Theorem~4. Therefore, the value of the tetrahedron coincides with the quantum $6j$ symbol, as expected.   \\

\section{Conclusions} \label{con}
\noindent
We have analyzed the Riemannian Einstein-Hilbert theory of gravity in $2+1D$, entailing $SU(2)$ internal symmetry, and shown that, when an additional cosmological constant term is considered, imposing constraints at the quantum level induces an effective recoupling theory that is the one proper of the $SU_q(2)$ quantum group. This amounts,  at the quantum level, to replacing the standard expressions for the amplitudes encoding elements of the recoupling theory of $SU(2)$ with elements of the recoupling theory of $SU_q(2)$. This has brought to verify the dynamical emergence of the Turaev-Viro model, as expected by comparison with the different perspective of quantization provided in \cite{WittenJonesPolinomials}. \\

Implementing the physical projector with cosmological constant, we have provided explicit computations of the physical inner product of two loop states, showing in detail the emergence of the deformed $SU_q(2)$ recoupling theory. We have further discussed how the physical projector implements the diffeomorphism invariance, and finally described the emergence of the Turaev-Viro model in the theory. \\

Instead of quantizing the reduced phase-space of the theory, we have shown here that the action of the curvature constraint at the quantum level induces the emergence of the effective recoupling theory, both at the level of the representation of the fundamental loop and at the level of the higher spin loops representations. Switching from the loop to the spin-network basis, adopting the very same symmetrization of representations, finally entails the effective equivalent expression for the intertwiners of the theory. This latter observation sheds light on the possible way to deform the internal  $SL(2, \mathbb{C})$ symmetry of Lorentzian theories of gravity in $3+1D$, providing a constructive argument based on a physical insight.

\acknowledgements 
\noindent
The authors acknowledge interesting discussions with Alejandro Perez and Carlo Rovelli that have brought to develop this analysis. NG was supported by the Natural Science Foundation of the Jiangsu Higher Education Institutions of China Programme grant No. 19KJB140018 and Xi'an Jiaotong-Liverpool University through grant No. REF-18-02-03. A.M.~wishes to acknowledge support by the Natural Science Foundation of China, through the grant No. 11875113, the Shanghai Municipality, through the grant No.~KBH1512299, and by Fudan University, through the grant No.~JJH1512105. 

\appendix

\end{document}